\documentclass[
  preprint,
  superscriptaddress,
  amsmath,
  amssymb,
  aps,
  prb,
  floatfix
]{revtex4-2}
\usepackage{graphicx}
\usepackage{dcolumn}
\usepackage{bm}
\usepackage{placeins}
\usepackage[
  colorlinks=true,
  citecolor=blue,
  linkcolor=blue,
  urlcolor=blue
]{hyperref}

\allowdisplaybreaks

\begin{document}

\title{Effect of uniform bond-charge order on longitudinal spin-susceptibility jump at a superconducting instability inside a magnetic phase}

\author{Sher Baz}
\affiliation{Department of Condensed Matter Physics, Graduate School of Science, Hokkaido University, Sapporo 060-0810, Japan}
\affiliation{Research Center for Materials Nanoarchitectonics (MANA), National Institute for Materials Science (NIMS), Tsukuba 305-0047, Japan}

\author{Hiroyuki Yamase}
\affiliation{Research Center for Materials Nanoarchitectonics (MANA), National Institute for Materials Science (NIMS), Tsukuba 305-0047, Japan}
\affiliation{Department of Condensed Matter Physics, Graduate School of Science, Hokkaido University, Sapporo 060-0810, Japan}
\date{\today}

\begin{abstract}
A previous work [H. Yamase and M. Zafur, Phys. Rev. B \textbf{103}, 224527 (2021)] showed that the longitudinal spin susceptibility exhibits a jump when the superconductivity develops continuously inside a magnetic phase. That analysis employed a minimal model containing magnetic and superconducting interactions. However, in correlated-electron systems, both magnetic and bond-charge interactions originate from the same underlying spin-spin exchange interaction. We therefore extend the model by including the uniform bond-charge order to provide a more realistic description of the system. We find that the jump of the longitudinal spin susceptibility remains robust even in the presence of uniform bond-charge order, but the uniform bond-charge order quantitatively enhances the longitudinal spin susceptibility and the size of the jump at $T_c$.
\end{abstract}
\maketitle
\clearpage


\section{Introduction}
\label{sec:introduction}
In correlated-electron systems, magnetic order often occurs in the close proximity to superconductivity, and coexistence of the two orders has been reported in several materials, including multilayer cuprates~\cite{mukuda12,kunisada20} and iron-based superconductors~\cite{pratt09,liu11}. The study of spin susceptibility in such a coexistence phase is particularly interesting because it may provide information on whether magnetism and superconductivity coexist microscopically as one electronic state or arise from spatially separated phases~\cite{yamase21b}. In particular, the longitudinal spin susceptibility directly probes the magnetic response of a state, in which superconductivity develops in the presence of preexisting magnetic order. 

The calculation of the longitudinal spin susceptibility inside a magnetically ordered phase is much more delicate than in the disordered phase~\cite{kuboki17}.
A conventional calculation of the spin susceptibility in a magnetically ordered phase takes into account the quasiparticle states in the ordered phase and possible Umklapp processes when translational symmetry is broken. For the longitudinal spin susceptibility, however, special care is needed~\cite{kuboki17}. In the disordered phase, the electron density and chemical potential are quadratic functions of the magnetic field. Hence, there is no contribution to the linear response quantities. However, in the magnetically ordered phase, spin-rotational symmetry is broken, and various physical quantities such as the density and the chemical potential gain a linear term as a function of an external magnetic field $h$~\cite{kuboki17}. Because of this, the longitudinal spin susceptibility at fixed density, $\chi_n$, behaves differently from that at fixed chemical potential, $\chi_\mu$, in the magnetically ordered phase~\cite{kuboki17,yamase21b}.

This unusual feature of the magnetically ordered phase leads to a further consequence when the system exhibits a superconducting instability. When superconductivity develops continuously inside the magnetic phase, the superconducting order parameter $\Delta_0$ also acquires a linear response to the external magnetic field~\cite{yamase21b}. In particular, the derivative $\partial\Delta_0/\partial h$ becomes singular as the superconducting transition is approached from below, resulting in a finite jump of the longitudinal spin susceptibility at $T_c$. The mechanism of this jump is different from the well-known specific-heat jump in BCS theory and originates from the broken spin-rotational symmetry of the magnetic phase. This behavior was captured transparently within mean-field theory, without resorting to a conventional diagrammatic technique, which becomes considerably more complicated inside the ordered phase ~\cite{kuboki17}.

The previous analysis was restricted to magnetic and superconducting interactions. However, the bond-charge interaction is also naturally connected to the spin-spin interaction, from which both magnetic and bond-charge contributions can emerge~\cite{yamase21c,bejas12,sachdev13}. Bond-charge and related charge-order correlations have also been widely discussed experimentally in correlated-electron systems and are often closely intertwined with superconductivity~\cite{da-silva-neto15,da-silva-neto16,da-silva-neto18}. Such bond-charge tendencies have been discussed in several theoretical studies of correlated-electron systems~\cite{allais14,yamase15b,li17,yamase19}. More recently, spin and bond-charge responses have been studied on an equal footing within an effective model containing these interactions~\cite{zafur24}. For a more realistic description of the system, one should consider a possible role of bond-charge order. This raises a natural question: does the longitudinal spin-susceptibility jump remain robust when bond-charge order is included?

In this work, we study the longitudinal spin susceptibility in a coexistence phase of superconductivity and magnetism by considering an effective model containing magnetic, superconducting, and bond-charge interactions, with a particular focus on uniform bond-charge order for simplicity. We employ mean-field theory, which allows transparent calculations of the longitudinal spin susceptibility at both fixed density and fixed chemical potential~\cite{kuboki17,yamase21b} and, most importantly, captures the essential physics. We find that the uniform bond-charge interaction modifies the coexistence phase by suppressing magnetic order and enhancing superconductivity, thereby increasing the longitudinal spin susceptibility and the magnitude of its jump at $T_c$. The microscopic origin of the jump, however, remains unchanged~\cite{yamase21b}; the singular contribution is associated with the field-induced superconducting response $\partial\Delta_0/\partial h$ inside the magnetic phase.

The rest of this paper is organized as follows. In Sec.~II, we introduce the effective Hamiltonian containing magnetic, superconducting, and bond-charge interactions and present the corresponding mean-field formulation. In Sec.~III, we present the numerical results for the longitudinal spin susceptibility under fixed-density conditions and examine the effect of the uniform bond-charge order on the longitudinal spin susceptibility. In Sec.~IV, we discuss the obtained results.  Section~V summarizes the main conclusions. The fixed-chemical-potential formulation and additional derivations are given in the appendices.
\section{Model and Formalism}

We employ an effective square-lattice model with magnetic, singlet-pairing, and bond-charge interactions, which allows the magnetic and bond-charge interactions to be treated on an equal footing. The Hamiltonian is given by
\begin{align}
H=&-\sum_{ij\sigma} t_{ij} c^{\dagger}_{i\sigma}c_{j\sigma}+V_m\sum_{i\tau}\hat{m}_i\hat{m}_{i+\tau}+V_b\sum_{i\tau}\hat{\chi}^{\dagger}_{i\tau}\hat{\chi}_{i\tau}
\nonumber\\&+V_s\sum_{i\tau}\hat{\Delta}^{\dagger}_{i\tau}\hat{\Delta}_{i\tau}-\sum_i h_i\hat{m}_i .
\label{eq:Hamiltonian}
\end{align}
Here, $c^{\dagger}_{i\sigma}$ and $c_{i\sigma}$ are the creation and annihilation operators of an electron with spin $\sigma$ at site $i$, respectively. The index $i$ runs over the sites of a square lattice, and $\tau=x,y$ denotes the nearest-neighbor bond directions. The hopping amplitude $t_{ij}$ takes the values $t$, $t'$, and $t''$ for nearest-neighbor, next-nearest-neighbor, and third-nearest-neighbor hoppings, respectively, and is zero otherwise.

The first term in Eq.~(\ref{eq:Hamiltonian}) represents the kinetic energy of itinerant electrons. The second term describes the magnetic interaction between nearest-neighbor sites. The local magnetization operator is defined as
\begin{equation}
\hat{m}_i=\frac{1}{2}\left(c^{\dagger}_{i\uparrow}c_{i\uparrow}-c^{\dagger}_{i\downarrow}c_{i\downarrow}\right).
\label{eq:m_operator}
\end{equation}
When $V_m$ is chosen to be positive, this interaction favors antiferromagnetic order. The third term in Eq.~(\ref{eq:Hamiltonian}) is the bond-charge interaction. The bond-charge operator is given by
\begin{equation}
\hat{\chi}_{i\tau}=\sum_{\sigma}c^{\dagger}_{i\sigma}c_{i+\tau,\sigma}.
\label{eq:chi_operator}
\end{equation}
This operator describes the electronic amplitude on the bond connecting sites $i$ and $i+\tau$. This should be distinguished from an onsite charge density, since the latter measures the charge on a lattice site, whereas $\hat{\chi}_{i\tau}$ measures the electronic amplitude on the nearest-neighbor bond. The coupling constant $V_b<0$ controls the strength of this bond-charge interaction.
The fourth term in Eq.~(\ref{eq:Hamiltonian}) describes the nearest-neighbor Cooper interaction. The corresponding pairing operator is defined by
\begin{equation}
\hat{\Delta}_{i\tau}=c_{i\uparrow}c_{i+\tau\downarrow}-c_{i\downarrow}c_{i+\tau\uparrow}.
\label{eq:Delta_operator}
\end{equation}
We take $V_s<0$, so that this interaction favors superconductivity. The last term in Eq.~(\ref{eq:Hamiltonian}) is the Zeeman coupling to an infinitesimal magnetic field $h_i$. Since the magnetic order considered here is N\'{e}el order, the longitudinal field is chosen to have the same modulation vector as the ordered moment $h_i=h e^{i{\bf Q}\cdot{\bf r}_i}$ with ordering wavevector ${\bf Q}=(\pi,\pi)$.

The coupling constants $V_m$, $V_b$, and $V_s$ characterize the magnetic, bond-charge, and singlet-pairing interactions, respectively. We take $V_m=J$ and $V_b=V_s=-3J/8$. This choice is motivated by the slave-boson mean-field theory of the \textit{t-J} model~\cite{lee06}, which also leads to the same interaction as Eq.~(\ref{eq:Hamiltonian}), although no strong correlation effect is present in our model.
To capture the essential physics of the longitudinal response in the ordered phase, we use mean-field theory, which is equivalent to the random phase approximation (RPA) and provides a transparent treatment of the additional field-induced terms~\cite{kuboki17,yamase21b}.

Hence we decouple the interaction terms in Eq.~(\ref{eq:Hamiltonian}) by introducing the mean fields,
\begin{equation}
\chi_{\tau} =\left\langle\hat{\chi}_{i\tau}\right\rangle ,
\label{eq:chi_mean}
\end{equation}
\begin{equation}
m = \left\langle\hat{m}_i\right\rangle e^{-i{\bf Q}\cdot{\bf r}_i},
\label{eq:m_mean}
\end{equation}
\begin{equation}
\Delta_{\tau}=\left\langle \hat{\Delta}_{i\tau} \right\rangle .
\label{eq:Delta_mean}
\end{equation}
We assume that these mean fields are real and uniform. For the superconducting state, we can check that \(d\)-wave symmetry gives the lowest energy; therefore, we assume \(\Delta_x = -\Delta_y = \Delta_0\). In addition, the bond-charge interaction is taken to be uniform, \(\chi_x = \chi_y = \chi_b\).

After mean-field decoupling and Fourier transformation, the Hamiltonian is written as

\begin{equation}
H_{\rm MF}=\sum_{{\bf k}}^{\prime}\Psi^{\dagger}_{\bf k}\hat{M}_{\bf k}\Psi_{\bf k}-2N\left(V_b\chi_b^2-V_m m^2+V_s\Delta_0^2\right)
\label{eq:HMF}
\end{equation}
The prime on the momentum summation indicates that ${\bf k}$ is restricted to the magnetic Brillouin zone, $|k_x|+|k_y|\leq \pi$, while $N$ is the total number of lattice sites.
We have introduced the four-component Nambu spinor
\begin{equation}
\Psi^{\dagger}_{\bf k}=\left(c^{\dagger}_{{\bf k}\uparrow},c_{-{\bf k}\downarrow},c^{\dagger}_{{\bf k}+{\bf Q}\uparrow},c_{-{\bf k}+{\bf Q}\downarrow}\right).
\label{eq:spinor}
\end{equation}
In this basis, the mean-field matrix is given by
\begin{equation}
\hat{M}_{\bf k}=\begin{pmatrix}
\xi_{\bf k} & -\Delta_{\bf k} & -\bar{m} & 0\\
-\Delta_{\bf k} & -\xi_{\bf k} & 0 & -\bar{m}\\
-\bar{m} & 0 & \xi_{{\bf k}+{\bf Q}} & -\Delta_{{\bf k}+{\bf Q}}\\
0 & -\bar{m} & -\Delta_{{\bf k}+{\bf Q}} & -\xi_{{\bf k}+{\bf Q}}
\end{pmatrix}.
\label{eq:MF_matrix}
\end{equation}
Here, $\xi_{\mathbf{k}}$ denotes the electronic dispersion,
\begin{align}
\xi_{\mathbf{k}}=&-2\Big[(t-V_b\chi_b)(\cos k_x+\cos k_y)+2t'\cos k_x\cos k_y \nonumber\\&\qquad+t''(\cos 2k_x+\cos 2k_y)\Big]-\mu .
\label{eq:xik}
\end{align}
The effective magnetic field is $\bar{m}=2V_m m+h/2$, while the superconducting gap is $\Delta_{\mathbf{k}}=2V_s\Delta_0 d_{\mathbf{k}}$, with the $d$-wave form factor $d_{\mathbf{k}}=\cos k_x-\cos k_y$.

The matrix $\hat M_{\mathbf{k}}$ is diagonalized by the unitary transformation
$\widetilde{\Psi}_{\mathbf{k}}=\hat U_{\mathbf{k}}^\dagger\Psi_{\mathbf{k}}$. The mean-field Hamiltonian can then be written as
\begin{align}
H_{\rm MF}
=&\sum_{\mathbf{k}}'
\widetilde{\Psi}_{\mathbf{k}}^{\dagger}
\begin{pmatrix}
\lambda_{\mathbf{k}}^{+} & 0 & 0 & 0\\
0 & -\lambda_{\mathbf{k}}^{+} & 0 & 0\\
0 & 0 & \lambda_{\mathbf{k}}^{-} & 0\\
0 & 0 & 0 & -\lambda_{\mathbf{k}}^{-}
\end{pmatrix}
\widetilde{\Psi}_{\mathbf{k}}
\nonumber\\
&-2N\left(V_b\chi_b^2-V_m m^2+V_s\Delta_0^2\right)-\mu N .
\label{eq:HMF_diagonal}
\end{align}
Diagonalization of the matrix $\hat{M}_{\bf k}$ yields the quasiparticle energies $\pm\lambda_{\bf k}^{+}$ and $\pm\lambda_{\bf k}^{-}$, where $\lambda_{\bf k}^{\pm}=\sqrt{(\eta_{\bf k}^{\pm})^2+\Delta_{\bf k}^2}$, $\eta_{\bf k}^{\pm}=\xi_{\bf k}^{+}\pm D_{\bf k}$, $D_{\bf k}=\sqrt{(\xi_{\bf k}^{-})^2+\bar{m}^2}$, and $\xi_{\bf k}^{\pm}=(\xi_{\bf k}\pm\xi_{{\bf k}+{\bf Q}})/2$. 

The grand-canonical potential per lattice site is obtained from $\Omega=-(\beta N)^{-1}\ln Z$, where $Z={\rm Tr}\,e^{-\beta H_{\rm MF}}$ and $\beta$ is the inverse of temperature $T$, i.e., $\beta=1/T$.
\begin{align}
\Omega=&-\frac{1}{\beta N}\sum_{\bf k}^{\prime}\Bigg[\ln\left(1+e^{-\beta\lambda_{\bf k}^{+}}\right)
+\ln\left(1+e^{\beta\lambda_{\bf k}^{+}}\right)\nonumber\\
&\hspace{1.2cm}+\ln\left(1+e^{-\beta\lambda_{\bf k}^{-}}\right)+\ln\left(1+e^{\beta\lambda_{\bf k}^{-}}\right)\Bigg]\nonumber\\
&-2\left(V_b\chi_b^2-V_m m^2+V_s\Delta_0^2\right)-\mu.
\label{eq:free_energy}
\end{align}
The self-consistency equations are obtained by minimizing the grand-canonical potential of Eq.~(\ref{eq:free_energy}): $\partial\Omega/\partial\chi_b=0$, $\partial\Omega/\partial\Delta_0=0$, and $\partial\Omega/\partial m=0$. The electron density is calculated from, $n=-\partial\Omega/\partial\mu$.
The resulting self-consistency equations are
\begin{align}
\chi_b&=-\frac{1}{2N}\sum_{\mathbf{k}}'
\frac{\xi^{-}_{\mathbf{k}}}{D_{\mathbf{k}}}s_{\mathbf{k}}\left[\frac{\eta^+_{\mathbf{k}}}{\lambda^+_{\mathbf{k}}}\tanh\left(\frac{\lambda^+_{\mathbf{k}}}{2T}\right)-\frac{\eta^-_{\mathbf{k}}}{\lambda^-_{\mathbf{k}}}
\tanh\left(\frac{\lambda^-_{\mathbf{k}}}{2T}\right)\right],
\label{eq:chi_gap}
\\
\Delta_0
&=
-\frac{1}{2N}\sum_{\mathbf{k}}'d_{\mathbf{k}}\left[\frac{\Delta_{\mathbf{k}}}{\lambda^+_{\mathbf{k}}}\tanh\left(\frac{\lambda^+_{\mathbf{k}}}{2T}\right)
+\frac{\Delta_{\mathbf{k}}}{\lambda^-_{\mathbf{k}}}\tanh\left(\frac{\lambda^-_{\mathbf{k}}}{2T}\right)\right],
\label{eq:Delta_gap}
\\
m
&=\frac{1}{2N}
\sum_{\mathbf{k}}'\frac{\bar{m}}{D_{\mathbf{k}}}\left[\frac{\eta^+_{\mathbf{k}}}{\lambda^+_{\mathbf{k}}}
\tanh\left(\frac{\lambda^+_{\mathbf{k}}}{2T}\right)
-\frac{\eta^-_{\mathbf{k}}}{\lambda^-_{\mathbf{k}}}\tanh\left(\frac{\lambda^-_{\mathbf{k}}}{2T}\right)\right],
\label{eq:m_gap}
\\
n
&=
1-\frac{1}{N}\sum_{\mathbf{k}}'\left[\frac{\eta^+_{\mathbf{k}}}{\lambda^+_{\mathbf{k}}}\tanh\left(\frac{\lambda^+_{\mathbf{k}}}{2T}\right)+\frac{\eta^-_{\mathbf{k}}}{\lambda^-_{\mathbf{k}}}\tanh\left(\frac{\lambda^-_{\mathbf{k}}}{2T}\right)\right],
\label{eq:n_eq}
\end{align}
where $s_{\bf k}=\cos k_x+\cos k_y$.

The longitudinal spin susceptibility is defined as
\begin{equation}
\chi_{n(\mu)}=\lim_{h\to 0}\left.\frac{\partial m}{\partial h}\right|_{n(\mu)},
\end{equation}
where the subscript specifies whether the electron density $n$ or the chemical potential $\mu$ is kept fixed. In a magnetically ordered phase, these two susceptibilities are not equivalent in general~\cite{yamase21b,kuboki17}. In the following, we formulate the longitudinal spin susceptibility $\chi_n$ under the condition of the fixed electron density. The corresponding formulation for the fixed chemical potential, denoted by $\chi_\mu$, is presented in Appendix~\ref{app:fixed_muu}.

At the fixed electron density, the chemical potential varies with the applied longitudinal field and acquires a linear term in $h$ inside the magnetic phase. In the coexistence phase, the superconducting order parameter also responds linearly to the field $h$, giving rise to the field-induced response $\partial\Delta_0/\partial h$, as discussed previously~\cite{yamase21b}. In the present model, the uniform bond-charge interaction introduces the additional self-consistent mean field $\chi_b$. Therefore, inside the magnetic phase, $\chi_b$ can also acquire a linear response to the field. Consequently, the relevant field-induced responses at the fixed density are $\partial\mu/\partial h$, $\partial\chi_b/\partial h$, $\partial\Delta_0/\partial h$, and $\partial\bar m/\partial h$.

The susceptibility at the fixed electron density $\chi_n$ is calculated by differentiating the self-consistency equations Eqs.~(\ref{eq:chi_gap})--(\ref{eq:n_eq}) with respect to the field $h$ and then taking the limit $h\rightarrow 0$ for the fixed-density condition. We obtain
\begin{equation}
\begin{pmatrix}
a_{11} & a_{12} & a_{13} & a_{14} \\
a_{21} & a_{22} & a_{23} & a_{24} \\
a_{31} & a_{32} & a_{33} & a_{34} \\
a_{41} & a_{42} & a_{43} & a_{44}
\end{pmatrix}
\begin{pmatrix}
\dfrac{\partial \mu}{\partial h} \\
\dfrac{\partial \chi_b}{\partial h} \\
\dfrac{\partial \Delta_0}{\partial h} \\
\dfrac{\partial \bar{m}}{\partial h}
\end{pmatrix}
=
\begin{pmatrix}
0 \\
0 \\
0 \\
\chi_n
\end{pmatrix}.
\label{eq:linear_response_n}
\end{equation}
The explicit expressions of the matrix elements $a_{ij}$ are listed in Appendix~\ref{app:matrix}. We denote the $4\times4$ matrix in Eq.~(\ref{eq:linear_response_n}) by $\hat{A}$, and define $\tilde{a}_{ij}$ as the $(i,j)$ cofactor of $\hat{A}$. Solving Eq.~(\ref{eq:linear_response_n}) gives $(\partial\bar{m}/\partial h)=(\tilde{a}_{44}/\det\hat{A})\chi_n$. On the other hand, from $\bar{m}=2V_m m+h/2$, we have $\partial\bar{m}/\partial h=2V_m\chi_n+1/2$. Combining these relations, we obtain
\begin{equation}
\begin{pmatrix}
\partial\mu/\partial h \\
\partial\chi_b/\partial h \\
\partial\Delta_0/\partial h \\
\chi_n
\end{pmatrix}
=
\frac{\chi_n^0}{1-4V_m\chi_n^0}
\begin{pmatrix}
\tilde{a}_{41}/\det\hat{A} \\
\tilde{a}_{42}/\det\hat{A} \\
\tilde{a}_{43}/\det\hat{A} \\
1
\end{pmatrix}.
\label{eq:response_solution_n}
\end{equation}
Here $\chi_n^0=\det\hat{A}/(2\tilde{a}_{44})$, and $\tilde{a}_{44}$ denotes the
cofactor of the $(4,4)$ element of $\hat{A}$.
\section{Results}
\label{sec:results}

We present the numerical results using the magnetic interaction strength $V_m$ as the unit of energy and take $t/V_m=0.7$, $t'/t=-0.14$, $t''/t=0.07$, and $V_b/V_m=V_s/V_m=-3/8$. The hopping parameters are chosen to yield a representative band structure for correlated electron systems~\cite{damascelli03,armitage10}. For $V_b=0$, the model reduces to the magnetic-superconducting model studied previously for the longitudinal spin-susceptibility jump~\cite{yamase21b}.

\subsection{Phase diagrams }

\begin{figure*}[t]
\centering
\includegraphics[width=0.47\textwidth]{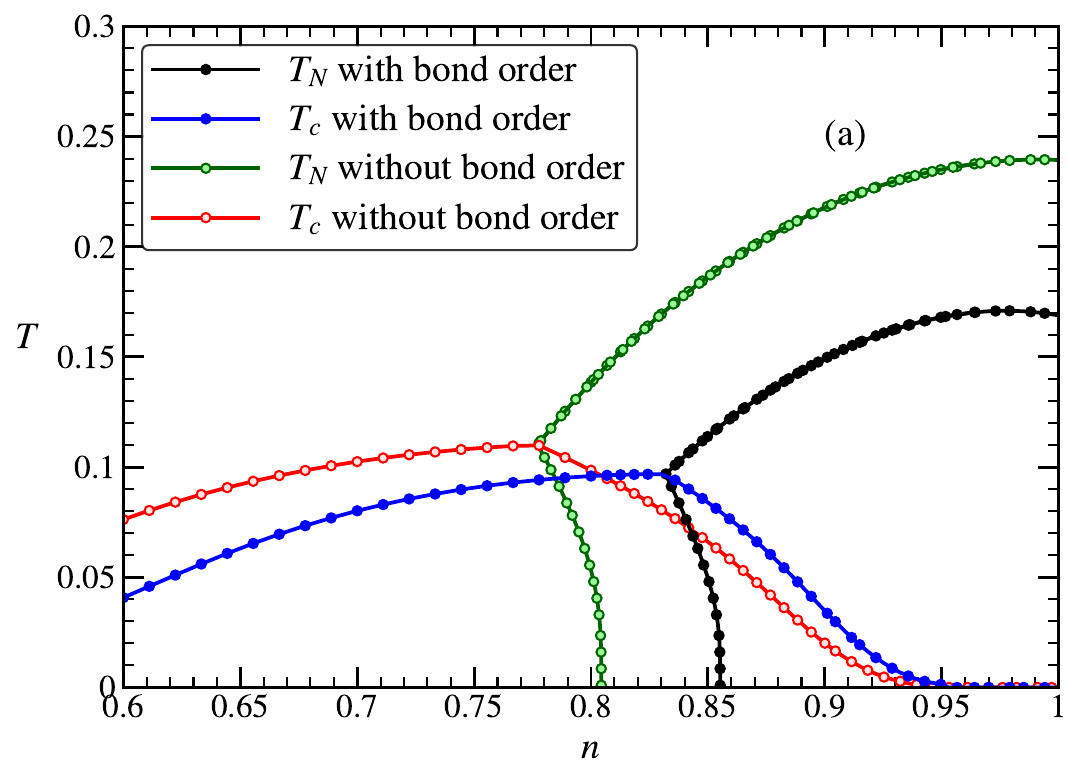}
\hfill
\includegraphics[width=0.47\textwidth]{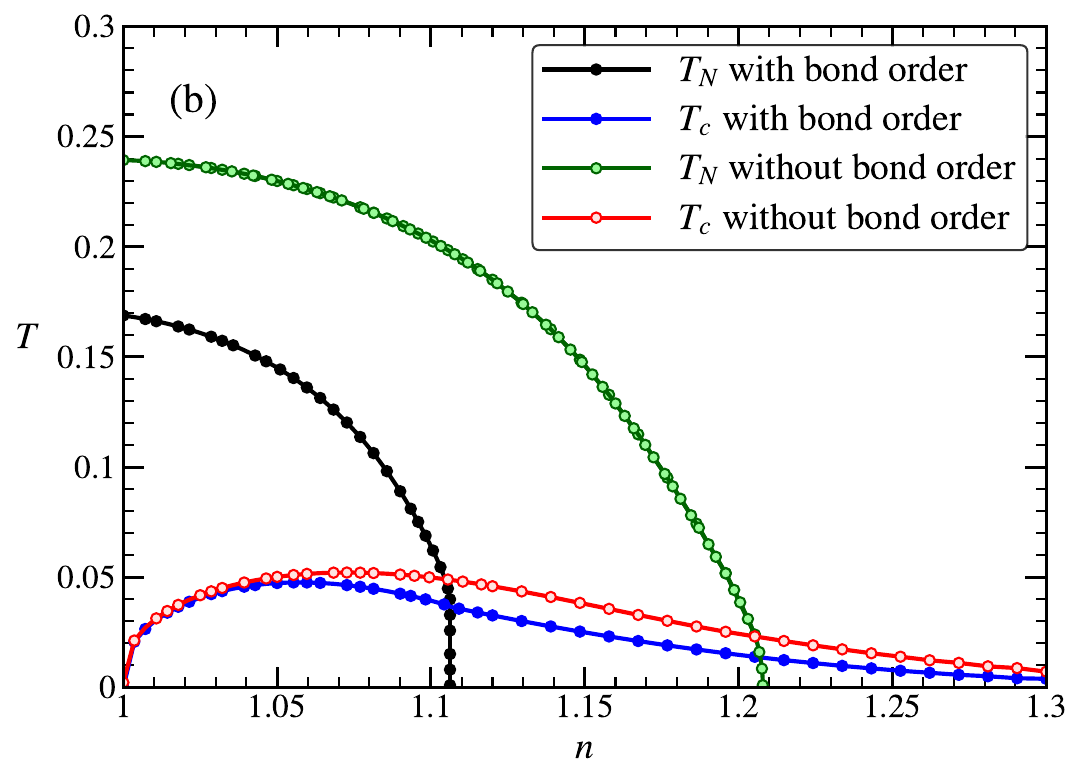}
\caption{Phase diagrams in the plane of electron density $n$ and temperature $T$ on (a) the hole-doped side and (b) the electron-doped side. The phase boundaries are obtained by assuming continuous phase transitions. The black and blue solid lines denote $T_N$ and $T_c$, respectively, in the presence of the uniform bond-charge interaction. The green and red lines show the corresponding results for the magnetic-superconducting model without the uniform bond-charge order.}
\label{fig:phase_diagrams}
\end{figure*}

Fig.~\ref{fig:phase_diagrams} shows the phase diagrams obtained from the self-consistency equations in the plane of electron density $n$ and temperature $T$. We compare the results with and without the uniform bond-charge interaction on both the hole- and electron-doped sides. The high-temperature region is the normal phase. Below $T_N$, magnetic order develops, and $d$-wave superconductivity appears below $T_c$. When both order parameters are finite, the magnetic and superconducting orders coexist. The overall structure and the pronounced asymmetry between the hole- and electron-doped sides are qualitatively similar to those found in the previous magnetic-superconducting model~\cite{yamase21b}. On the hole-doped side, as shown in Fig.~\ref{fig:phase_diagrams}(a), the inclusion of the uniform bond-charge order leads to a pronounced suppression of magnetism relative to the corresponding magnetic-superconducting model. In particular, the magnetic transition temperature $T_N$ is reduced, and the magnetic phase is confined to a narrower density range closer to half-filling. The superconducting transition temperature $T_c$ is also modified, although its variation is considerably weaker. The same qualitative trend is found on the electron-doped side in Fig.~\ref{fig:phase_diagrams}(b). The uniform bond-charge order markedly lowers $T_N$ and shifts the magnetic phase toward half-filling. In contrast, the corresponding change in $T_c$ is comparatively small. Consequently, the magnetic-superconducting coexistence region becomes narrower, while the pure superconducting phase extends. Thus, uniform bond-charge order changes the phase diagram primarily through its strong suppression of magnetic order.

\subsection{Fixed-density result}
\label{density}
\begin{figure*}[t]
\centering
\includegraphics[width=0.48\textwidth]{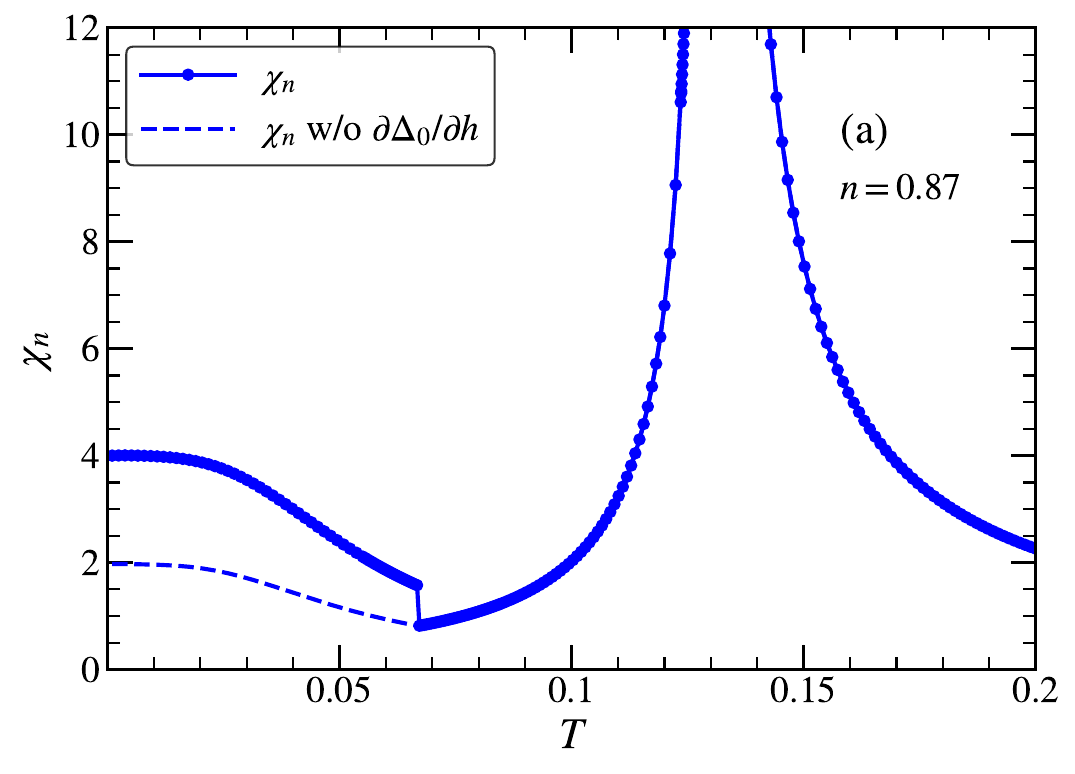}
\hfill
\includegraphics[width=0.48\textwidth]{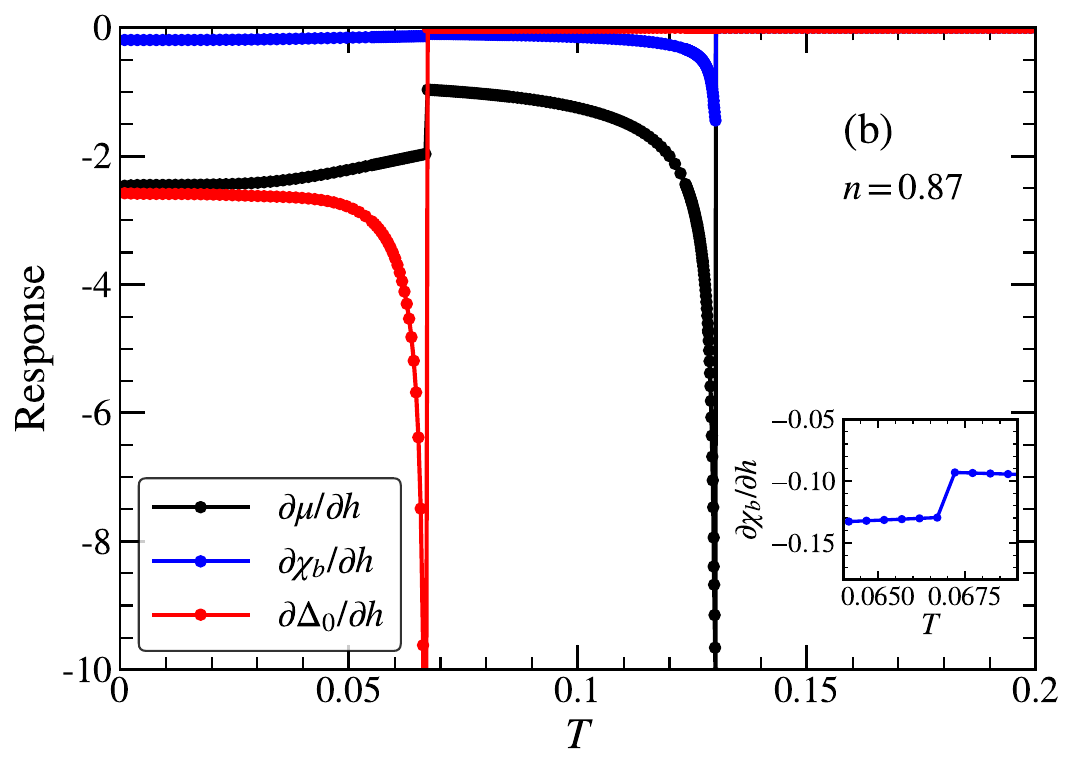}
\caption{Temperature dependence at fixed density $n=0.87$. (a) Longitudinal spin susceptibility $\chi_n$. Magnetic and superconducting instabilities occur at $T_N=0.132$ and $T_c=0.067$, respectively. The abbreviation ``w/o'' in (a) means ``without". The dashed curve shows the result obtained without the contribution from $\partial\Delta_0/\partial h$. (b) Temperature dependence of field-induced responses $\partial\mu/\partial h$, $\partial\chi_b/\partial h$, and $\partial\Delta_0/\partial h$. The inset magnifies the jump in $\partial\chi_b/\partial h$ at $T_c$.}
\label{fig:fixed_density_bond}
\end{figure*}

Figure~\ref{fig:fixed_density_bond}(a) shows the temperature dependence of the longitudinal spin susceptibility $\chi_n$ for the fixed density $n=0.87$. The system is in the normal metallic phase at high temperatures. With decreasing temperature, $\chi_n$ is enhanced and diverges at the magnetic transition temperature $T_N=0.132$, showing the instability toward the magnetic phase. Below $T_N$, the magnetic order develops, and the susceptibility is rapidly suppressed. With further decreasing temperature, the system undergoes a superconducting instability at $T_c=0.067$ via a continuous phase transition, below which magnetism and superconductivity coexist. Despite the continuous onset of superconductivity, $\chi_n$ exhibits a finite jump at $T_c$, similar to that discussed previously~\cite{yamase21b}.

Fig.~\ref{fig:fixed_density_bond}(b) shows the temperature dependence of the induced linear responses $\partial\mu/\partial h$, $\partial\chi_b/\partial h$, and $\partial\Delta_0/\partial h$. These responses arise because spin-rotational symmetry is broken in the magnetic phase~\cite{yamase21b,kuboki17}. The responses $\partial\mu/\partial h$ and $\partial\chi_b/\partial h$ diverge at $T_N$ and become finite below $ T_N$, whereas $\partial\Delta_0/\partial h$ diverges at $T_c$  and becomes finite only in the coexistence phase. This singular superconducting response produces the jump of $\chi_n$ at $T_c$, as established previously for the magnetic-superconducting model~\cite{yamase21b}. Similarly, both $\partial\mu/\partial h$ and $\partial\chi_b/\partial h$ also exhibit finite jumps at $T_c$, as shown in Fig.~\ref{fig:fixed_density_bond}(b). When the contribution from $\partial\Delta_0/\partial h$ is omitted, the jump disappears, and only a cusp remains, as shown by the dashed curve in Fig.~\ref{fig:fixed_density_bond}(a). Thus, the singularity of the superconducting response is clearly responsible for the jump of the longitudinal spin susceptibility.

To understand the divergences of $\partial\mu/\partial h$ and $\partial\chi_b/\partial h$ at $T_N$ and $\partial\Delta_0/\partial h$ at $T_c$ observed in Fig.~\ref{fig:fixed_density_bond}(b), we focus first on $T\rightarrow T_c^{-}$, where $\Delta_0$ is infinitesimally small. Expanding the matrix elements in Eq.~(\ref{eq:linear_response_n}) in powers of $\Delta_0$, we find $a_{13},a_{23},a_{31},a_{32},a_{34},a_{43}\propto\Delta_0$ and $a_{33}\propto\Delta_0^2$. Consequently, $\det\hat{A}$, $\tilde{a}_{41}$, $\tilde{a}_{42}$, and $\tilde{a}_{44}$ are proportional to $\Delta_0^2$, whereas $\tilde{a}_{43}\propto\Delta_0$. From the leading $\Delta_0$ dependence of the matrix elements, one obtains $\det\hat{A}\approx \det\hat{A}^{\prime\prime}\Delta_0^2$, $\tilde{a}_{41}\approx \tilde{a}_{41}^{\prime\prime}\Delta_0^2$, $\tilde{a}_{42}\approx \tilde{a}_{42}^{\prime\prime}\Delta_0^2$, $\tilde{a}_{43}\approx \tilde{a}_{43}^{\prime}\Delta_0$, and $\tilde{a}_{44}\approx \tilde{a}_{44}^{\prime\prime}\Delta_0^2$. The coefficients carrying a single or double prime are finite at $T_c$, with the single (double) prime referring to the coefficient of the term linear (quadratic) in $\Delta_0$. Substituting these leading behaviors into Eq.~(\ref{eq:response_solution_n}), we obtain, in the limit $T\rightarrow T_c^{-}$,

\begin{equation}
\begin{pmatrix}
\partial\mu/\partial h \\
\partial\chi_b/\partial h \\
\partial\Delta_0/\partial h \\
\chi_n
\end{pmatrix}
=\frac{\chi_n^{0-}}{1-4V_m\chi_n^{0-}}
\begin{pmatrix}
\tilde{a}_{41}^{\prime\prime}/\det\hat{A}^{\prime\prime} \\
\tilde{a}_{42}^{\prime\prime}/\det\hat{A}^{\prime\prime} \\
(\tilde{a}_{43}^{\prime}/\det\hat{A}^{\prime\prime})\Delta_0^{-1} \\
1
\end{pmatrix},
\label{eq:response_below_tc}
\end{equation}
Thus, the longitudinal spin susceptibility at $T=T_c^{-}$ is given by
\begin{equation}
\chi_n(T_c^-)=\frac{\chi_n^{0-}}{1-4V_m\chi_n^{0-}},
\label{eq:chi0_below}
\end{equation}
where
\begin{equation}
\chi_n^{0-}=\frac{1}{2}\frac{\det\hat{A}^{\prime\prime}}{\tilde{a}_{44}^{\prime\prime}}.
\end{equation}

Eq.~(\ref{eq:response_below_tc}) shows that $\chi_n$, $\partial\mu/\partial h$, and $\partial\chi_b/\partial h$ are finite as $T\to T_c^-$, whereas $\partial\Delta_0/\partial h\propto\Delta_0^{-1}$. Since $\Delta_0\propto(T_c-T)^{1/2}$ in mean-field theory, it follows that
\begin{equation}
\frac{\partial\Delta_0}{\partial h}\propto\Delta_0^{-1}\propto(T_c-T)^{-1/2},
\qquad T\rightarrow T_c^{-}.
\label{eq:delta_response_divergence}
\end{equation}
Hence, among the field-induced responses, only $\partial\Delta_0/\partial h$ becomes singular at $T_c$, consistent with Fig.~\ref{fig:fixed_density_bond}(b). For $T\to T_c^{-}$, the field-induced responses are determined by the full $4\times4$ response matrix $\hat A$. Although the common factors of $\Delta_0^2$ cancel in $\chi_n^{0-}$, the resulting finite value still contains the effect of superconductivity through the coefficients of the leading quadratic terms in $\Delta_0$.

For  $T\to T_c^{+}$, $\Delta_0=0$ and $\partial\Delta_0/\partial h$ is absent, so the $4\times4$ response equation reduces to
\begin{equation}
\begin{pmatrix}
c_{11} & c_{12} & c_{13} \\
c_{21} & c_{22} & c_{23} \\
c_{31} & c_{32} & c_{33}
\end{pmatrix}
\begin{pmatrix}
\partial\mu/\partial h \\[2pt]
\partial\chi_b/\partial h \\[2pt]
\partial\bar{m}/\partial h
\end{pmatrix}
=
\begin{pmatrix}
0 \\[2pt]
0 \\[2pt]
\chi_n
\end{pmatrix},
\label{eq:response_above_tc}
\end{equation}
where $c_{11}=a_{11}$, $c_{12}=a_{12}$, $c_{13}=a_{14}$, $c_{21}=a_{21}$, $c_{22}=a_{22}$, $c_{23}=a_{24}$, $c_{31}=a_{41}$, $c_{32}=a_{42}$, and $c_{33}=a_{44}$ in Appendix~\ref{app:matrix}. We denote the $3\times3$ matrix in Eq.~(\ref{eq:response_above_tc}) by $\hat{C}$, and let $\tilde{c}_{ij}$ denote the $(i,j)$ cofactor of $\hat{C}$. The reduced matrix equation yields $\partial\bar{m}/\partial h=(\tilde{c}_{33}/\det\hat{C})\chi_n$. Combining this relation with $\partial\bar{m}/\partial h=2V_m\chi_n+1/2$, the longitudinal spin susceptibility at $T=T_c^{+}$ is expressed as 
\begin{equation}
\chi_n(T_c^+)=\frac{\chi_n^{0+}}{1-4V_m\chi_n^{0+}},
\label{eq:chi0_above_tc}
\end{equation}
where
\begin{equation}
\chi_n^{0+}=\frac{1}{2}\frac{\det\hat{C}}{\tilde{c}_{33}}.
\end{equation}
Here, $\chi_n^{0+}$ is determined by the quantities in the nonsuperconducting magnetic and bond-ordered phase. Eqs.~(\ref{eq:chi0_below}) and~(\ref{eq:chi0_above_tc}) describe the longitudinal spin susceptibility below and above the superconducting transition $T_c$, respectively, and are obtained from different response matrices. Hence, $\chi_n^{0-}$ and $\chi_n^{0+}$ are generally different; this is a reason why the longitudinal spin susceptibility $\chi_n$ shows a finite jump at $T_c$. Similarly, $\partial\mu/\partial h$ and $\partial\chi_b/\partial h$ also take different values below and above $T_c$, leading to finite jumps in both responses at $T_c$.

On the other hand, at the magnetic transition $T_N$, the responses $\partial\mu/\partial h$ and $\partial\chi_b/\partial h$ are governed by the leading $m$ dependence of the reduced response matrix. As $T\to T_N^{-}$, the matrix elements $c_{13}$, $c_{23}$, $c_{31}$, and $c_{32}$ are linear in the magnetic order parameter, whereas the other elements are finite. Consequently, $\det\hat{C}$ is finite, while the cofactors entering the first two components of the response equation satisfy $\tilde{c}_{31},\tilde{c}_{32}=O(m)$. Therefore, simple algebra shows that $\partial\mu/\partial h$ and $\partial\chi_b/\partial h$ scale as $m\chi_n$. Since $\chi_n\propto(T_N-T)^{-\gamma}$ and $m\propto(T_N-T)^{\beta}$, these responses behave as $(T_N-T)^{\beta-\gamma}$. For the mean-field exponents $\gamma=1$ and $\beta=1/2$, it follows that
\begin{equation}
\frac{\partial\mu}{\partial h},\,\frac{\partial\chi_b}{\partial h}\propto (T_N-T)^{-1/2},\qquad T\to T_N^- .
\end{equation}
This behavior accounts for the divergences of $\partial\mu/\partial h$ and $\partial\chi_b/\partial h$ at $T_N^-$, as shown in Fig.~\ref{fig:fixed_density_bond}(b). 

\subsection{Effect of the uniform bond-charge order on the jump of the spin susceptibility}
\label{subsec:bond_effect}

\begin{figure}[t]
\centering
\includegraphics[width=0.48\linewidth]{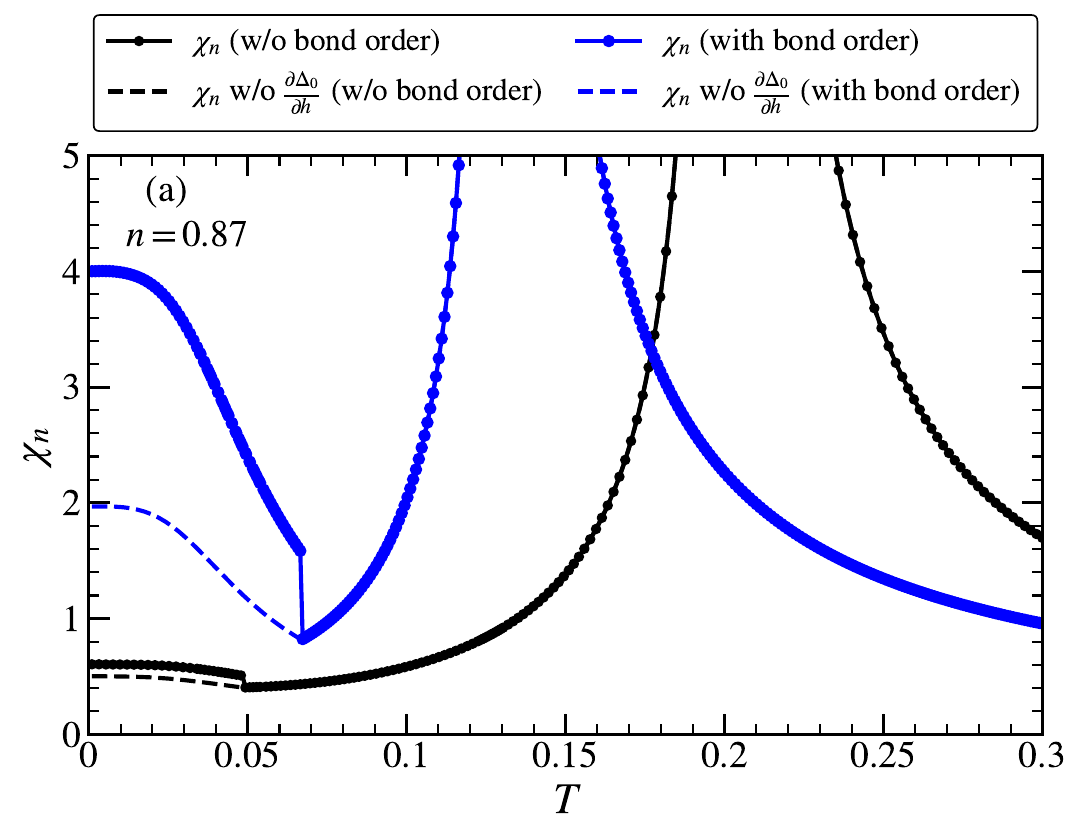}
\hfill
\includegraphics[width=0.48\linewidth]{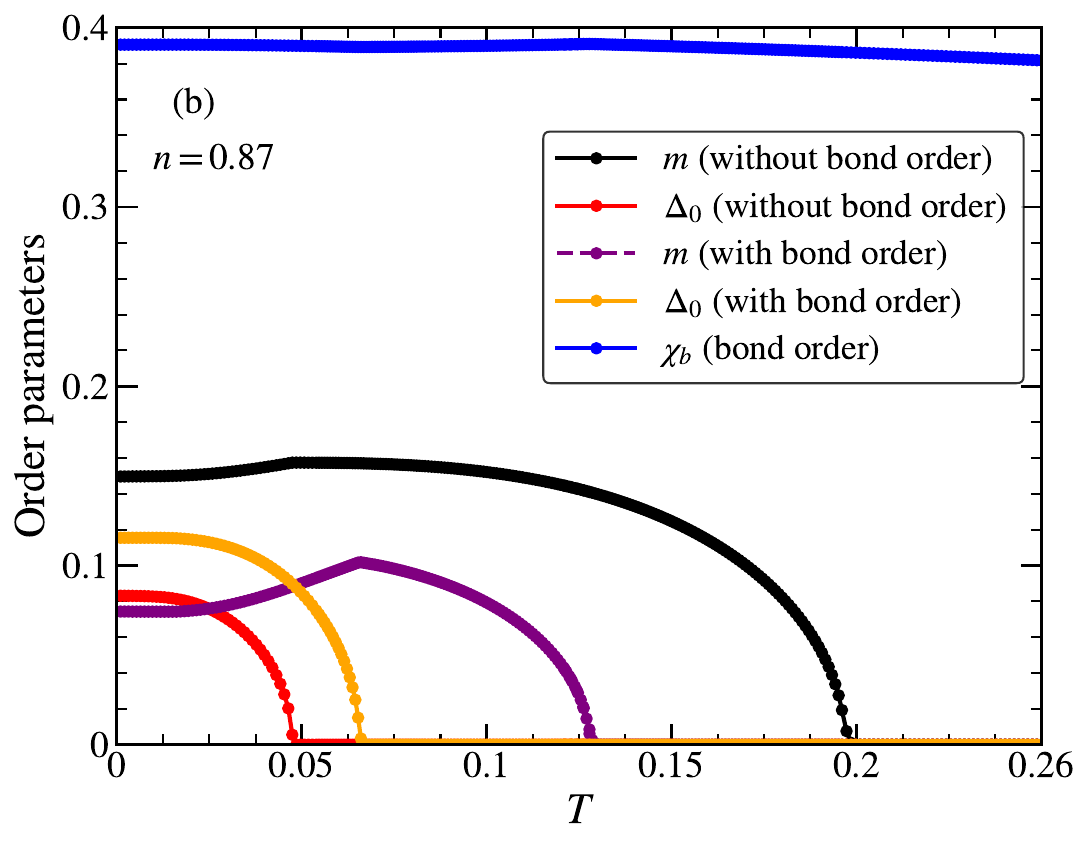}
\caption{Comparison between the models with and without the uniform bond-charge order at the fixed density $n=0.87$. (a) Temperature dependence of the longitudinal spin susceptibility, $\chi_n$. The abbreviation ``w/o'' in (a) means ``without". Solid and dashed curves show the results with and without the contribution from $\partial\Delta_0/\partial h$, respectively. (b) Temperature dependence of the order parameters $m$, $\Delta_0$, and $\chi_b$.}
\label{fig:fixed_density_comparison}
\end{figure}

To clarify the role of the uniform bond-charge order, Fig.~\ref{fig:fixed_density_comparison} compares the fixed-density results with and without the uniform bond-charge order at $n=0.87$. The uniform bond-charge order produces substantial quantitative changes. The magnetic transition temperature is reduced from approximately $T_N=0.20$ to $0.132$, whereas the superconducting transition temperature increases from approximately $T_c=0.05$ to $0.067$. The longitudinal spin susceptibility is also strongly modified, with a pronounced enhancement in the low-temperature coexistence phase and a considerably larger jump at $T_c$. The corresponding order parameters are shown in Fig.~\ref{fig:fixed_density_comparison}(b). In the presence of the uniform bond-charge order, the magnetic order parameter $m$ is suppressed, whereas the superconducting order parameter $\Delta_0$ is enhanced. Below $T_c$, the development of $\Delta_0$ further suppresses $m$, demonstrating the competition between magnetic order and superconductivity in the coexistence phase. The uniform bond-charge mean field $\chi_b$ varies only weakly with temperature.

The enhancement of $\chi_n$ can be understood from the uniform bond-charge-induced modification of the coexistence phase. Magnetic order suppresses the longitudinal spin susceptibility, while superconductivity weakens the magnetic order below $T_c$. Hence, the longitudinal spin response in the coexistence phase is enhanced. Consequently, both $\chi_n$ and the magnitude of its jump at $T_c$ are increased. When the contribution from $\partial\Delta_0/\partial h$ is omitted, the jump disappears in both models, as shown by the dashed curves in Fig.~\ref{fig:fixed_density_comparison}(a).

Using $\chi_n(T_c^-)$ and $\chi_n(T_c^+)$ from Eqs.~(\ref{eq:chi0_below}) and~(\ref{eq:chi0_above_tc}), respectively, we quantify the magnitude of the susceptibility jump by introducing the normalized jump
\begin{equation}
\frac{\Delta\chi_n}{\chi_n(T_c+\delta T)}
=\frac{\chi_n(T_c-\delta T)-\chi_n(T_c+\delta T)}{\chi_n(T_c+\delta T)},
\label{eq:normalized_jump}
\end{equation}
Here, we take $\delta T=0.00001$, and $\chi_n(T_c^-)$ and $\chi_n(T_c^+)$ denote the susceptibilities just below and above $T_c$, respectively. This dimensionless quantity allows a direct comparison of the jump at different electron densities and between the models with and without the uniform bond-charge order.
\begin{figure}[t]
\centering
\includegraphics[width=0.48\textwidth]{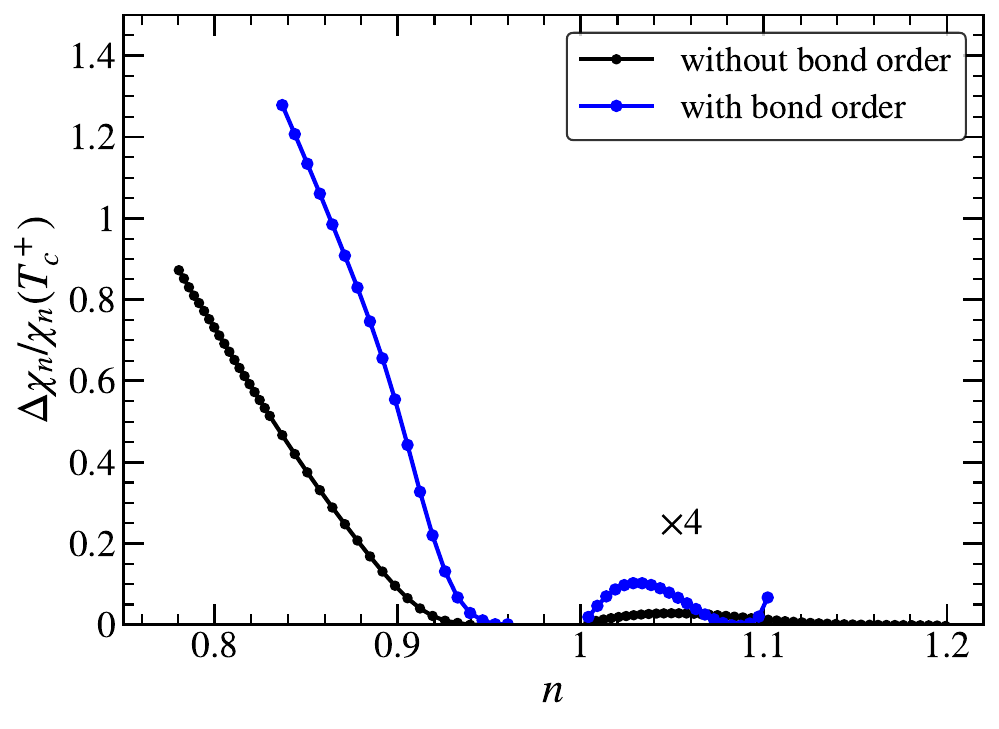}
\caption{Electron density dependence of the normalized jump $\Delta\chi_n/\chi_n(T_c^+)$ evaluated along the superconducting transition line inside the magnetic phase. The black curve denotes the magnetic-superconducting model without the uniform bond-charge order, whereas the blue curve denotes the model including the uniform bond-charge order. We search for the superconducting instability down to $T=0.0003$ for $0.95 \leq n \leq 1$ but do not find it; consequently, the normalized susceptibility jump is not defined in that density range. The electron-doped jump values are multiplied by a factor of four for clarity.}
\label{fig:jump_density}
\end{figure}
Fig.~\ref{fig:jump_density} shows the density dependence of the normalized jump along the superconducting transition line inside the magnetic phase. In both models, the magnitude of the jump depends strongly on electron density and is substantially larger on the hole-doped side than on the electron-doped side. This electron-hole asymmetry is discussed in Ref.~\cite{yamase21b,zafur24}. Because of the finite $t'$ and $t''$, the Fermi pockets are formed in different regions of momentum space on the two sides of half-filling. In the magnetic phase, the Fermi pockets are formed around $(\pm\pi/2,\pm\pi/2)$ and $(\pm\pi/2,\mp\pi/2)$ on the hole-doped side, and around $(\pm\pi,0)$ and $(0,\pm\pi)$ on the electron-doped side. Consequently, the distinct Fermi-pocket structures on the hole- and electron-doped sides lead to the pronounced asymmetry in the jump between the two doping regions.

The comparison between the two curves shows that the uniform bond-charge order enhances the normalized jump on both sides of half-filling. The enhancement is especially pronounced on the hole-doped side, where the jump is already sizable without the uniform bond-charge order and becomes substantially larger when the bond order is included. On the electron-doped side, the normalized jump is much smaller than on the hole-doped side; therefore, the electron-doped jump values in Fig.~\ref{fig:jump_density} are multiplied by a factor of four for clarity. With increasing density from half-filling, the normalized jump first increases and then decreases to a very small value. In the presence of the uniform bond-charge order, the normalized jump increases again near the termination of the coexistence phase around $n\approx1.1$. In this region, the magnetic denominator $1-4V_m\chi_n^{0-}$ becomes very small, while the susceptibility difference associated with the jump slightly increases again. These two effects lead to the upturn of the normalized jump. In contrast, without bond order, the corresponding susceptibility difference remains very small near the end of the coexistence phase, and no comparable upturn appears, although the denominator $1-4V_m\chi_n^{0-}$ is small. Therefore, the uniform bond-charge order makes the longitudinal susceptibility jump more pronounced over a substantial portion of the magnetic-superconducting coexistence phase.

\section{Discussion}
The comparison in Fig.~\ref{fig:fixed_density_comparison}(a) shows that the longitudinal spin susceptibility exhibits qualitatively similar behavior with and without the uniform bond-charge order. In particular, a finite jump appears at $T_c$ in both cases, and the susceptibility is enhanced inside the coexistence phase. This similarity indicates that the origin of the jump is the same as that clarified in Ref.~\cite{yamase21b}, where the jump was shown to rely on two general features: the emergence of $\partial\Delta_0/\partial h$ below $T_c$ inside the magnetic phase and its singular behavior at $T_c$ when superconductivity develops through a continuous phase transition. Thus, the susceptibility jump is robust against the inclusion of the uniform bond-charge order, and the underlying mechanism of the jump is the same.

Although the qualitative behavior is similar in the two models, the quantitative differences become pronounced in the coexistence phase. Figure~\ref{fig:fixed_density_comparison}(b) shows that the uniform bond-charge order changes the balance between the magnetic and superconducting orders, making $m$ and $\Delta_0$ comparable. This is important because the magnitude of the susceptibility jump is closely related to the competition between the two orders. As discussed in Ref.~\cite{yamase21b}, the jump tends to become pronounced when the magnetic order is substantially suppressed by the development of superconductivity. In the present model, the uniform bond-charge mean field $\chi_b$ renormalizes the quasiparticle dispersion through the effective nearest-neighbor hopping $t_{\rm eff}=t-V_b\chi_b$, where $V_b<0$, thereby suppressing the magnetic phase and modifying the magnetic-superconducting coexistence region. The uniform bond-charge order therefore enhances the jump of $\chi_n$  by changing the competition between magnetism and superconductivity in the coexistence phase. The inclusion of this channel is important for a realistic description of the system.

The enhancement of the susceptibility jump by the uniform bond-charge order is not restricted to a particular electron density but persists over a broad region of the magnetic-superconducting coexistence phase. In particular, the normalized jump is substantially larger on the hole-doped side than on the electron-doped side. This electron-hole asymmetry originates from the particle-hole asymmetry, as discussed in Ref.~\cite{yamase21b}.

The robustness of the susceptibility jump is further confirmed by the fixed-chemical-potential results presented in Appendix~\ref{app:fixed_muu}. The qualitative behavior remains largely similar to the fixed-density case, while the jump at $T_c$ is more pronounced for the representative fixed-chemical-potential case. The distinction between $\chi_n$ and $\chi_\mu$ in the magnetic phase and the persistence of the jump under both thermodynamic conditions were discussed previously in Refs.~\cite{kuboki17,yamase21b}. The fixed-chemical-potential case is relevant when one analyzes multilayer or multiband systems in terms of an effective single-layer or single-band model, provided that charge transfer between the layers or the bands likely occurs.

The results also strengthen the possible experimental relevance of the longitudinal spin-susceptibility jump. As discussed in Ref.~\cite{yamase21b}, a jump at the superconducting transition provides a thermodynamic signature of the microscopic coexistence of magnetism and superconductivity. The uniform bond-charge order makes the susceptibility jump more pronounced and therefore more favorable for experimental observation.

The present results are obtained within an effective mean-field treatment, and effects beyond this approximation may change quantitative aspects of the phase diagram and the magnitude of the susceptibility jump. In particular, stronger correlation effects and fluctuations may modify the balance between magnetic order, superconductivity, and the uniform bond-charge order. Nevertheless, the present results suggest that the persistence of the jump does not depend sensitively on these quantitative details, provided that superconductivity develops continuously inside the magnetic phase. 
\section{Conclusions}
\label{sec:conclusion}
We have studied the jump of the longitudinal spin susceptibility at a superconducting instability inside a magnetic phase by including the uniform bond-charge interaction in the magnetic-superconducting model. We find that the jump at $T_c$ remains robust in the presence of the uniform bond-charge order, while its microscopic origin remains unchanged and is governed by the field-induced response $\partial\Delta_0/\partial h$. The uniform bond-charge order modifies the magnetic-superconducting coexistence phase by suppressing the magnetic order and enhancing superconductivity, resulting in a larger longitudinal spin susceptibility and a more pronounced jump at $T_c$. The enhancement depends substantially on the carrier density and becomes more pronounced on the hole-doped side. These results show that the susceptibility jump is robust beyond the minimal magnetic-superconducting model and serve as a thermodynamic signature of microscopic coexistence between magnetism and superconductivity.
\begin{acknowledgments}
This work was supported by the World Premier International Center Initiative (WPI), MEXT, Japan.
\end{acknowledgments}
\appendix 
\section{Matrix elements for a fixed density}
\label{app:matrix}
The matrix elements $a_{ij}$ in Eq.~(\ref{eq:linear_response_n}) are obtained by differentiating the self-consistency equations Eqs.~(\ref{eq:chi_gap})--(\ref{eq:n_eq}) with respect to the external field $h$. They are given by
\begin{align}
a_{11}&=\frac{1}{N}\sum_{\bf k}'(P_{\bf k}^{+}+P_{\bf k}^{-}),\\
a_{12}&=-\frac{2V_b}{N}\sum_{\bf k}'\frac{\xi_{\bf k}^{-}s_{\bf k}}{D_{\bf k}}(P_{\bf k}^{+}-P_{\bf k}^{-}),\\
a_{13}&=-\frac{2V_s}{N}\sum_{\bf k}'\Delta_{\bf k}d_{\bf k}(Q_{\bf k}^{+}+Q_{\bf k}^{-}),\\
a_{14}&=-\frac{\bar m}{N}\sum_{\bf k}'\frac{1}{D_{\bf k}}(P_{\bf k}^{+}-P_{\bf k}^{-}),\\
a_{21}&=-\frac{a_{12}}{4V_b},\\
a_{22}&=-\frac{V_b}{N}\sum_{\bf k}'\Bigg[
\frac{\bar m^2s_{\bf k}^2}{D_{\bf k}^3}(R_{\bf k}^{+}-R_{\bf k}^{-})+\frac{(\xi_{\bf k}^{-})^2s_{\bf k}^2}{D_{\bf k}^2}
(P_{\bf k}^{+}+P_{\bf k}^{-})
\Bigg]-1,\\
a_{23}&=-\frac{V_b}{N}\sum_{\bf k}'\frac{\xi_{\bf k}^{-}\Delta_{\bf k}s_{\bf k}d_{\bf k}}{D_{\bf k}}(Q_{\bf k}^{+}-Q_{\bf k}^{-}),\\
a_{24}&=\frac{1}{2N}\sum_{\bf k}'\frac{\xi_{\bf k}^{-}\bar m s_{\bf k}}{D_{\bf k}^2}\Bigg[\frac{R_{\bf k}^{+}-R_{\bf k}^{-}}{D_{\bf k}}-(P_{\bf k}^{+}+P_{\bf k}^{-})
\Bigg],\\
a_{31}&=-\frac{a_{13}}{4V_s},\\
a_{32}&=a_{23},\\
a_{33}&=-\frac{V_s}{N}\sum_{\bf k}'d_{\bf k}^2(S_{\bf k}^{+}+S_{\bf k}^{-})-1,\\
a_{34}&=-\frac{1}{2N}\sum_{\bf k}'\frac{\bar m\Delta_{\bf k}d_{\bf k}}{D_{\bf k}}(Q_{\bf k}^{+}-Q_{\bf k}^{-}),\\
a_{41}&=\frac{a_{14}}{2},\\
a_{42}&=-2V_ba_{24},\\
a_{43}&=-2V_sa_{34},\\
a_{44}&=\frac{1}{2N}\sum_{\bf k}'\Bigg[
\frac{(\xi_{\bf k}^{-})^2}{D_{\bf k}^3}(R_{\bf k}^{+}-R_{\bf k}^{-})+\frac{\bar m^2}{D_{\bf k}^2}(P_{\bf k}^{+}+P_{\bf k}^{-})
\Bigg],
\end{align}
where we introduced
\begin{align}
P_{\bf k}^{\pm}
&=\frac{1}{\lambda_{\bf k}^{\pm}}\tanh\frac{\lambda_{\bf k}^{\pm}}{2T}+\frac{(\eta_{\bf k}^{\pm})^2}{\lambda_{\bf k}^{\pm}}g_{\bf k}^{\pm},
\\
Q_{\bf k}^{\pm}&=\frac{\eta_{\bf k}^{\pm}}{\lambda_{\bf k}^{\pm}}g_{\bf k}^{\pm},
\\
R_{\bf k}^{\pm}&=\frac{\eta_{\bf k}^{\pm}}{\lambda_{\bf k}^{\pm}}\tanh\frac{\lambda_{\bf k}^{\pm}}{2T},
\\
S_{\bf k}^{\pm}&=\frac{1}{\lambda_{\bf k}^{\pm}}\tanh\frac{\lambda_{\bf k}^{\pm}}{2T}+\frac{\Delta_{\bf k}^{2}}{\lambda_{\bf k}^{\pm}}g_{\bf k}^{\pm},
\end{align}
\begin{equation}
g_{\bf k}^{\pm}=-\frac{1}{(\lambda_{\bf k}^{\pm})^2}\tanh\frac{\lambda_{\bf k}^{\pm}}{2T}+\frac{1}{2T\lambda_{\bf k}^{\pm}}\frac{1}{\cosh^2\!\left(\dfrac{\lambda_{\bf k}^{\pm}}{2T}\right)}.
\end{equation}

\section{Results at fixed chemical potential}
\label{app:fixed_muu}

\begin{figure}[t]
\centering
\includegraphics[width=0.48\linewidth]{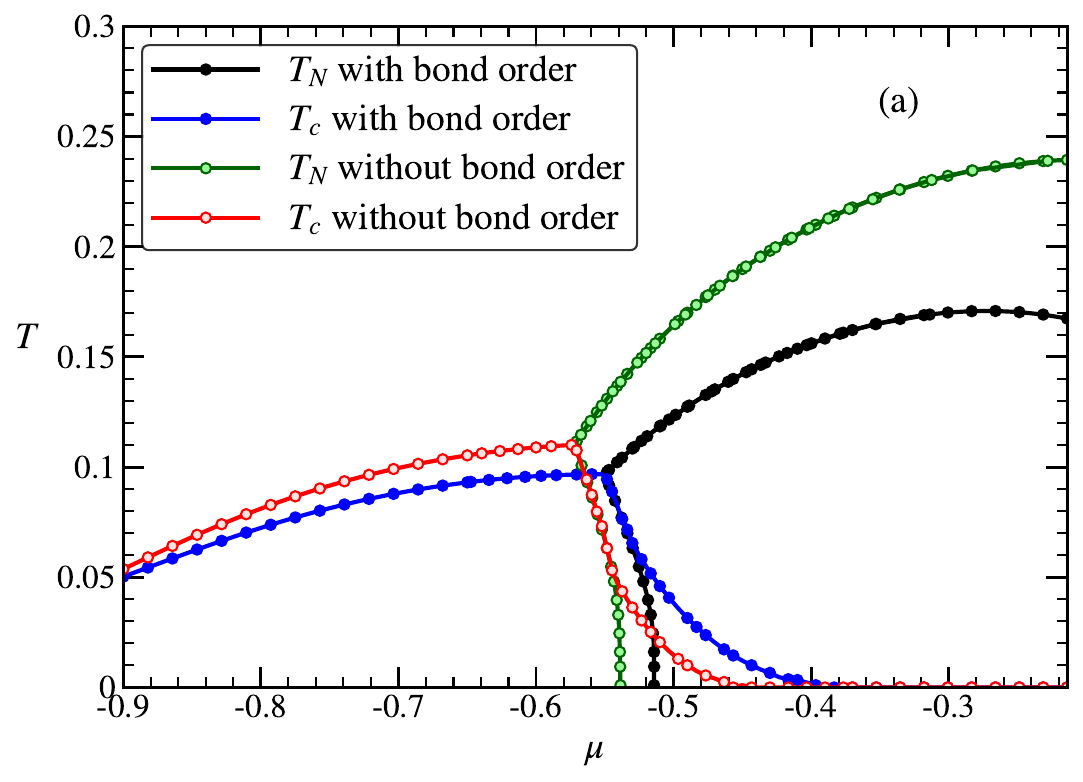}
\hfill
\includegraphics[width=0.48\linewidth]{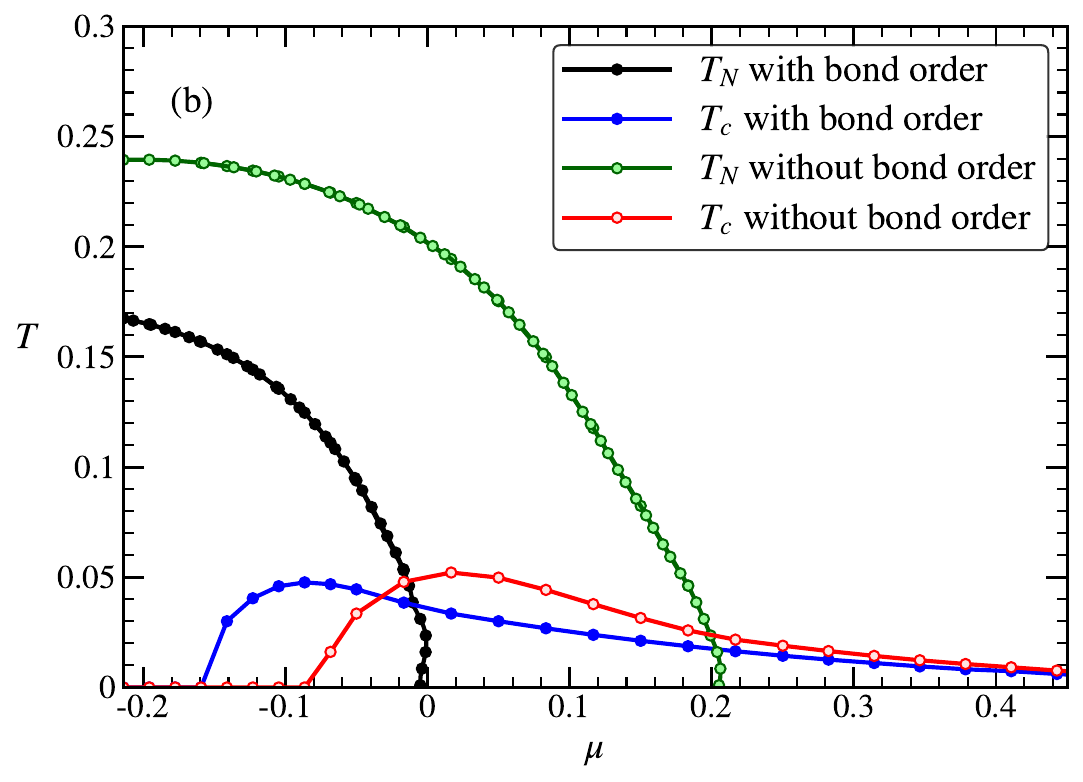}
\caption{Phase diagram in the plane of the chemical potential $\mu$ and  $T$ on (a) the hole-doped side and (b) the electron-doped side. The phase boundaries are obtained by assuming continuous phase transitions. The black and blue solid lines denote $T_N$ and $T_c$, respectively, in the presence of the uniform bond-charge order. The green and red lines show the corresponding results for the magnetic-superconducting model without the uniform bond-charge order.}
\label{fig:phase_mu}
\end{figure}

The longitudinal spin response depends on the thermodynamic constraint in the magnetic phase~\cite{kuboki17}. We here consider the fixed-chemical-potential case. We first present the phase diagram. Figure~\ref{fig:phase_mu} shows the phase diagram obtained at fixed chemical potential in the $\mu$--$T$ plane. The hole-doped and electron-doped sides are presented in Figs.~\ref{fig:phase_mu}(a) and \ref{fig:phase_mu}(b), respectively.

On the hole-doped side $T_N$ is reduced over the entire magnetic region, whereas the change in $T_c$ is comparatively modest. Consequently, the location and extent of the magnetic-superconducting coexistence phase are modified. A similar tendency is found on the electron-doped side. The uniform bond-charge order markedly lowers $T_N$ and restricts the magnetic phase to a narrower range of chemical potential. By contrast, $T_c$ is affected less strongly, although the superconducting phase boundary is shifted. As a result, the coexistence phase becomes narrower, while the pure superconducting phase extends into a larger region.

We consider the longitudinal spin susceptibility under the condition of fixed chemical potential. In this case $\mu$ is held constant, whereas the electron density is allowed to vary in response to the external field. The susceptibility is defined as $\chi_\mu=\left.\lim_{h\rightarrow0} \frac{\partial m}{\partial h} \right|_\mu $. Taking the derivative of the self-consistency equations Eqs.~(\ref{eq:chi_gap})--(\ref{eq:n_eq}) with respect to the field $h$ and then taking the limit $h\rightarrow 0$ for the fixed chemical potential $\mu$, we obtain
\label{app:fixed_mu_phase}
\begin{equation}
\begin{pmatrix}
b_{11} & b_{12} & b_{13} & b_{14} \\
b_{21} & b_{22} & b_{23} & b_{24} \\
b_{31} & b_{32} & b_{33} & b_{34} \\
b_{41} & b_{42} & b_{43} & b_{44}
\end{pmatrix}
\begin{pmatrix}
\dfrac{\partial n}{\partial h} \\
\dfrac{\partial \chi_b}{\partial h} \\
\dfrac{\partial \Delta_0}{\partial h} \\
\dfrac{\partial \bar{m}}{\partial h}
\end{pmatrix}
=
\begin{pmatrix}
0 \\
0 \\
0 \\
\chi_\mu
\end{pmatrix}.
\label{eq:linear_response_mu}
\end{equation}

The matrix in Eq.~(\ref{eq:linear_response_mu}) is denoted by $\hat{B}$. Its elements are the same as those of the fixed-density matrix $\hat{A}$, except for the first column, which reflects the replacement of $\partial\mu/\partial h$ by $\partial n/\partial h$. That is, $b_{ij}=a_{ij}$ except for $b_{11}=-1$ and     $b_{21}=b_{31}=b_{41}=0 $. Denoting the $(i,j)$ cofactor of $\hat{B}$ by $\tilde{b}_{ij}$, the solution is written as
\begin{equation}
\begin{pmatrix}
\partial n/\partial h \\
\partial\chi_b/\partial h \\
\partial\Delta_0/\partial h \\
\chi_\mu
\end{pmatrix}
=
\frac{\chi_\mu^0}{1-4V_m\chi_\mu^0}
\begin{pmatrix}
\tilde{b}_{41}/\det\hat{B} \\
\tilde{b}_{42}/\det\hat{B} \\
\tilde{b}_{43}/\det\hat{B} \\
1
\end{pmatrix},
\label{eq:response_solution_mu}
\end{equation}
where $\chi_\mu^0=\frac{\det\hat{B}}{2\tilde{b}_{44}}$.
\begin{figure*}[t]
\centering
\includegraphics[width=0.48\textwidth]{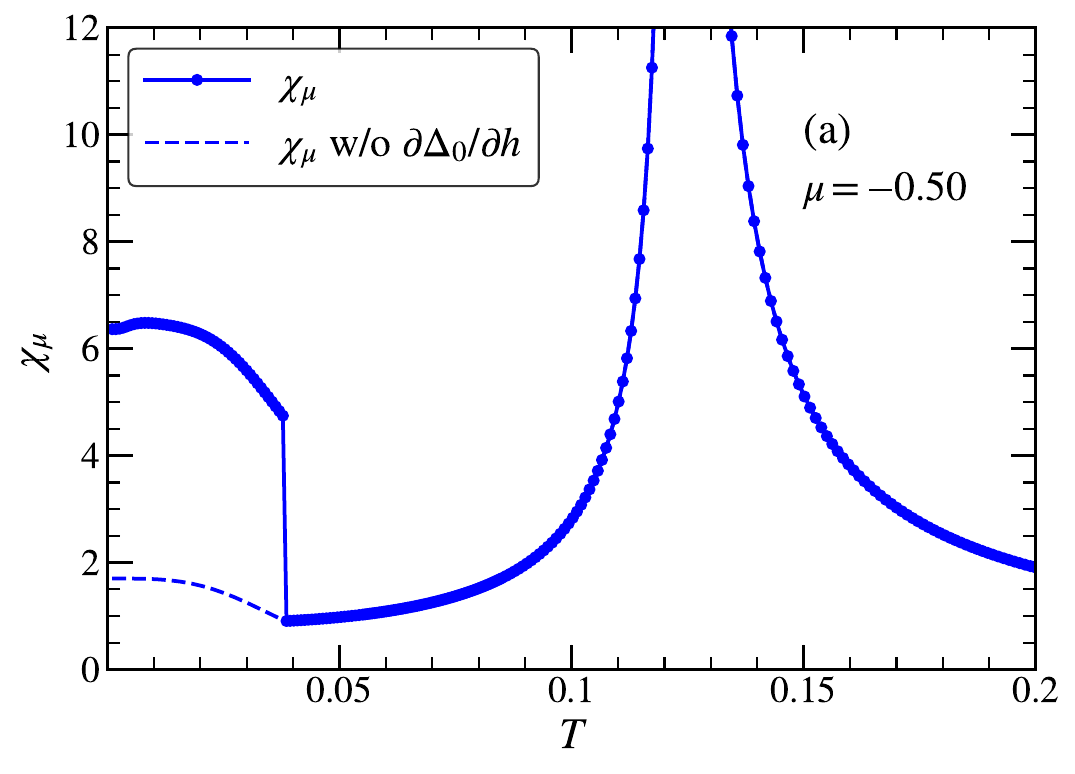}
\hfill
\includegraphics[width=0.48\textwidth]{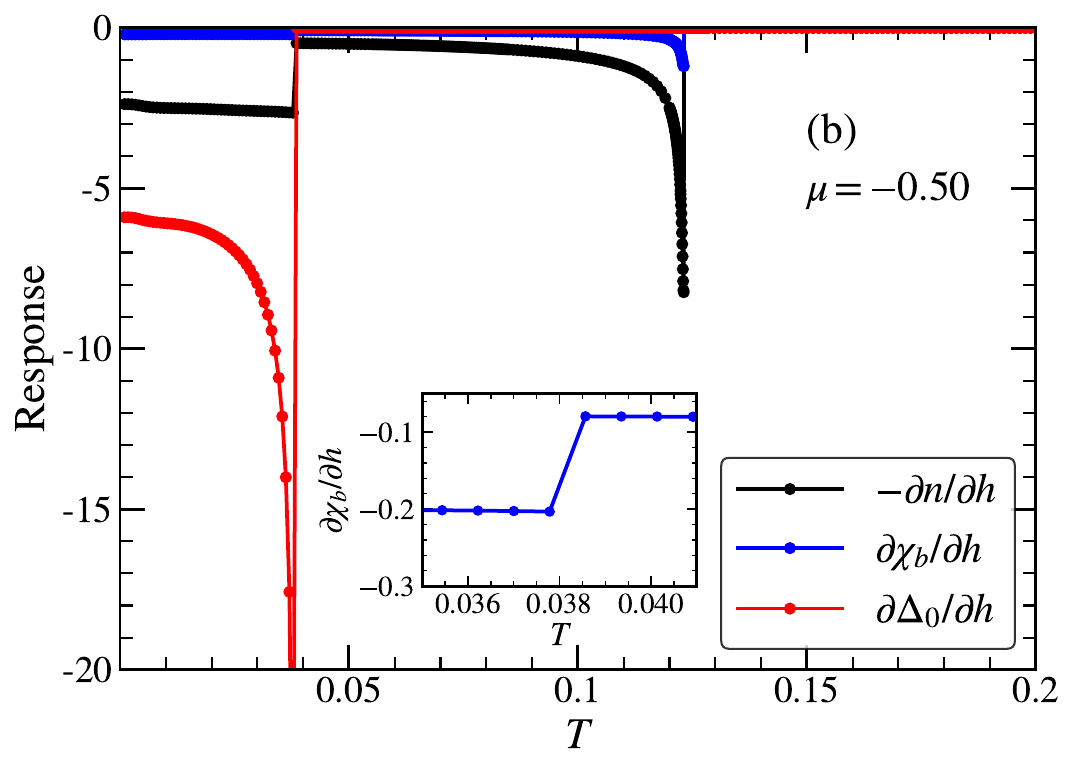}
\caption{Temperature dependence for a fixed chemical potential $\mu=-0.50$. (a) Longitudinal spin susceptibility $\chi_\mu$. The abbreviation ``w/o'' in (a) means ``without". The dashed curve shows the result obtained without the contribution from $\partial\Delta_0/\partial h$. (b) Temperature dependence of field-induced responses $-\partial n/\partial h$, $\partial\chi_b/\partial h$, and $\partial\Delta_0/\partial h$. Magnetic and superconducting instabilities occur at $T_N=0.123$ and $T_c=0.037$, respectively. The inset magnifies the jump in $\partial\chi_b/\partial h$ at $T_c$.}
\label{fig:fixed_mu_bond}
\end{figure*}

In Fig.~\ref{fig:fixed_mu_bond}(a), we show the temperature dependence of the longitudinal spin susceptibility at a fixed chemical potential $\mu=-0.50$. Although the electron density depends on temperature at fixed chemical potential, it remains in the vicinity of $n\approx 0.87$ over the temperature range considered. This density range closely corresponds to that considered in the fixed-density calculation at $n=0.87$ in Fig.~\ref{fig:fixed_density_bond}(a). Similar to the fixed-density case, $\chi_\mu$ diverges at $T_N=0.123$ and exhibits a jump at $T_c=0.037$. The jump is more pronounced for fixed chemical potential, consistent with the thermodynamic relation $\chi_\mu=\chi_n-\left.\frac{\partial n}{\partial h}\right|_{\mu}\left.\frac{\partial \mu}{\partial h}\right|_{n}=\chi_n+\left(\left.\frac{\partial n}{\partial h}\right|_{\mu}\right)^2\left.\frac{\partial \mu}{\partial n}\right|_{h}\geq\chi_n$ ~\cite{kuboki17,yamase21b}. The corresponding field-induced responses are shown in Fig.~\ref{fig:fixed_mu_bond}(b). In particular, $\partial\Delta_0/\partial h$ becomes singular at $T_c^{-}$, yielding the jump in $\chi_\mu$. Thus, the origin of the jump is the same as in the fixed-density case.

\begin{figure*}[t]
\centering
\includegraphics[width=0.48\textwidth]{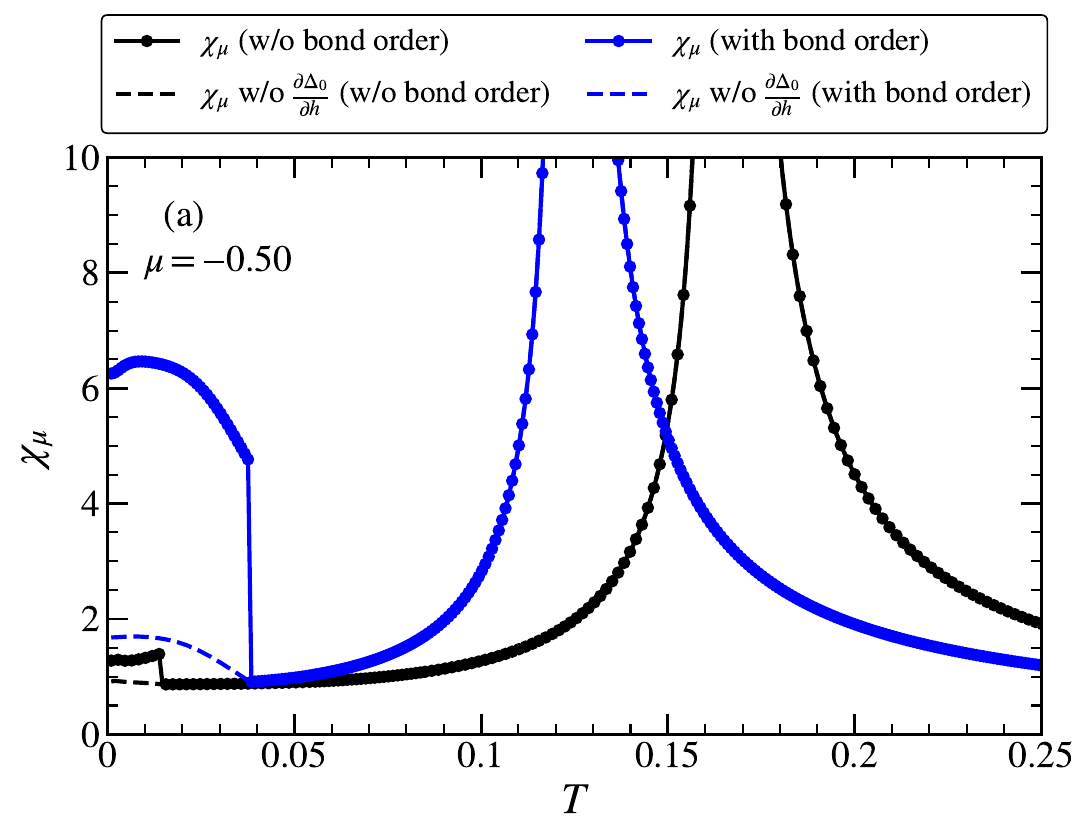}
\hfill
\includegraphics[width=0.48\textwidth]{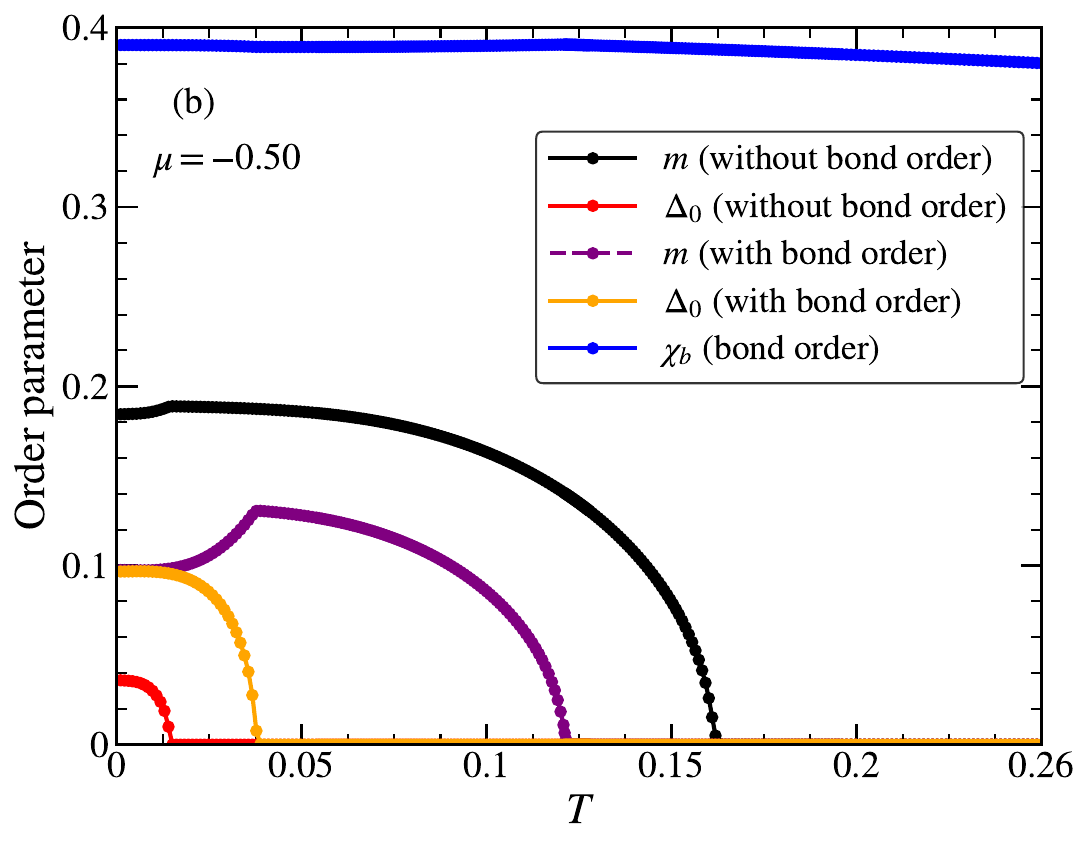}
\caption{Comparison between the models with and without the uniform bond-charge order at a fixed chemical potential $\mu=-0.50$. (a) Temperature dependence of the longitudinal spin susceptibility $\chi_\mu$. The abbreviation ``w/o'' in (a) means ``without". Solid and dashed curves denote results with and without the contribution from $\partial\Delta_0/\partial h$, respectively. (b) Temperature dependence of the order parameters $m$, $\Delta_0$, and $\chi_b$.}
\label{fig:fixed_mu_comparison}
\end{figure*}

Figure~\ref{fig:fixed_mu_comparison} compares the results at a fixed chemical potential $\mu=-0.50$ with and without the uniform bond-charge order. The qualitative behavior is similar to that found in the fixed-density case shown in Fig.~\ref{fig:fixed_density_comparison}: the uniform bond-charge order suppresses the magnetic order parameter $m$, enhances the superconducting order parameter $\Delta_0$, and increases the longitudinal spin susceptibility in the coexistence phase. However, the quantitative changes are more pronounced for the fixed-chemical-potential case. As shown in Fig.~\ref{fig:fixed_mu_comparison}(b), both $\Delta_0$ and $T_c$ are enhanced more strongly, accompanied by a larger enhancement of $\chi_\mu$ and its jump at $T_c$ in Fig.~\ref{fig:fixed_mu_comparison}(a). Despite these quantitative differences, the mechanism responsible for the jump is unchanged. In fact, when the contribution from $\partial\Delta_0/\partial h$ is omitted, the jump disappears in both models, as indicated by the dashed curves in Fig.~\ref{fig:fixed_mu_comparison}(a).

To quantify the jump at fixed chemical potential, we consider $\chi_\mu$ on the two sides of $T_c$. Following the same asymptotic analysis as for the fixed-density case, $\partial\Delta_0/\partial h$ diverges as $T\rightarrow T_c^{-}$, whereas $\chi_\mu$ remains finite. The  susceptibility below $T_c$ is $\chi_\mu(T_c^-)=\chi_\mu^{0-}/(1-4V_m\chi_\mu^{0-})$, with $\chi_\mu^{0-}=\det\hat{B}^{\prime\prime}/(2\tilde{b}_{44}^{\prime\prime})$. The quantities $\det\hat{B}^{\prime\prime}$ and $\tilde{b}_{ij}^{\prime\prime}$ denote the finite coefficients of the leading terms quadratic in $\Delta_0$. Above $T_c$, $\Delta_0=0$ and the superconducting response is absent, yielding $\chi_\mu(T_c^+)=\chi_\mu^{0+}/(1-4V_m\chi_\mu^{0+})$, with $\chi_\mu^{0+}=\det\hat{D}/(2\tilde{d}_{33})$. $\hat{D}$ denotes the reduced $3\times3$ coefficient matrix obtained from $\hat{B}$ in Eq.~(\ref{eq:linear_response_mu}) by removing the superconducting row and column, $\det\hat{D}$ denotes its determinant, and $\tilde{d}_{33}$ denotes its $(3,3)$ cofactor. Thus, $\chi_\mu$ generally takes different finite values on the two sides of $T_c$, leading to the jump at $T_c$. 

\begin{figure}[t]
\centering
\includegraphics[width=0.48\textwidth]{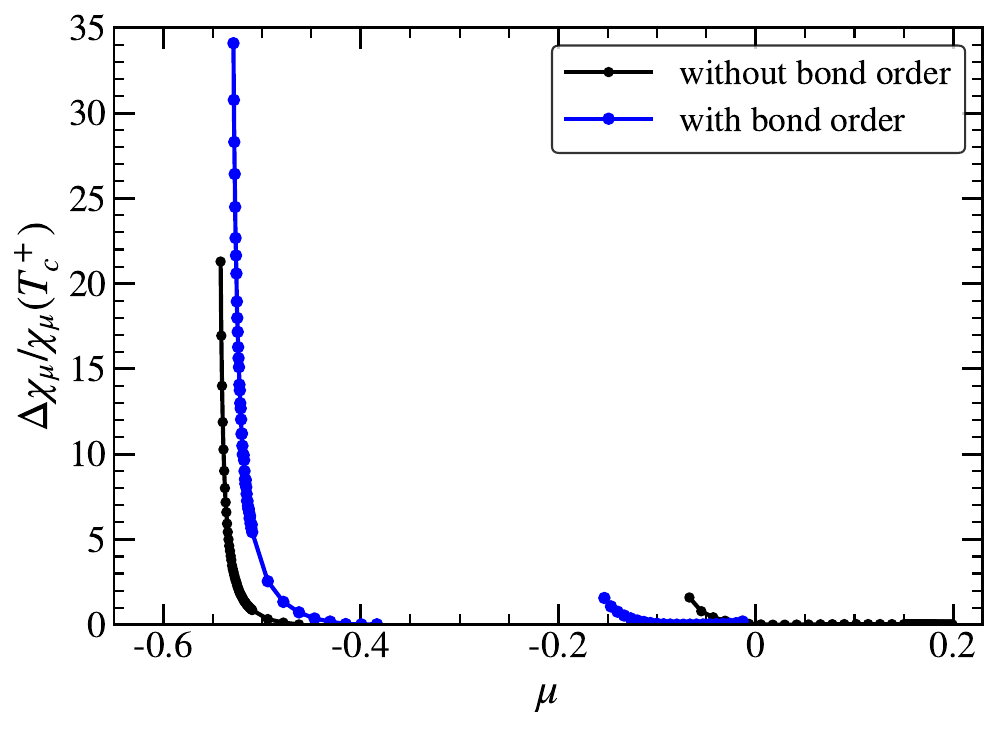}
\caption{Chemical-potential dependence of the normalized jump $\Delta\chi_\mu/\chi_\mu(T_c^+)$ along the superconducting transition line inside the magnetic phase. The black and blue curves denote the results without and with the uniform bond-charge order, respectively.}
\label{fig:jump_mu_comparison}
\end{figure}

In analogy with the fixed-density case, we characterize its magnitude by the normalized jump $\Delta\chi_\mu/\chi_\mu(T_c^+)$, where $\Delta\chi_\mu=\chi_\mu(T_c-\delta T)-\chi_\mu(T_c+\delta T)$ and $\delta T=0.00001$. Figure~\ref{fig:jump_mu_comparison} shows the chemical-potential dependence of the normalized jump along the superconducting transition line inside the magnetic phase. The overall behavior is similar to that obtained previously for the fixed-chemical-potential susceptibility~\cite{yamase21b}: the jump is generally more pronounced than in the fixed-density case and becomes strongly enhanced in the range $\mu=-0.55$ to $-0.50$. The large normalized values in this region should not be interpreted as a divergence of $\chi_\mu$ at $T_c$. As discussed in Ref.~\cite{yamase21b}, $\chi_\mu(T_c^+)$ becomes small while the absolute jump $\Delta\chi_\mu$ remains finite, leading to a large value of $\Delta\chi_\mu/\chi_\mu(T_c^+)$.  On the electron-doped side, the normalized jump behaves differently at fixed density and fixed chemical potential, as seen in Figs.~\ref{fig:jump_density} and ~\ref{fig:jump_mu_comparison}. 
This difference arises from the numerical details. At fixed density, $\chi_n(T_c^-)$ and $\chi_n(T_c^+)$ become nearly equal near half-filling, so that the absolute jump $\Delta\chi_n$ is strongly suppressed and the normalized jump remains small. In contrast, at fixed chemical potential, $\chi_\mu(T_c^+)$ decreases upon approaching the onset of the coexistence phase, whereas the absolute jump $\Delta\chi_\mu=\chi_\mu(T_c^-)-\chi_\mu(T_c^+)$ remains finite. The pronounced enhancement of $\Delta\chi_\mu/\chi_\mu(T_c^+)$ near the onset of the coexistence phase therefore originates mainly from the suppression of $\chi_\mu(T_c^+)$. As $\mu$ increases from the onset of the coexistence phase, $\chi_\mu(T_c^+)$ increases while $\Delta\chi_\mu$ decreases, and consequently the normalized jump becomes small over an intermediate range. The coexistence phase extends over a smaller range of $\mu$ with bond order because it terminates earlier, whereas without bond order it persists to larger $\mu$. Thus, the effect of the uniform bond-charge order is mainly quantitative.

\bibliography{main}

@article{allais14,
  title = {Density wave instabilities in a correlated two-dimensional metal},
  author = {Allais, Andrea and Bauer, Johannes and Sachdev, Subir},
  journal = {Phys. Rev. B},
  volume = {90},
  issue = {15},
  pages = {155114},
  numpages = {5},
  year = {2014},
  month = {Oct},
  publisher = {American Physical Society},
  optdoi = {10.1103/PhysRevB.90.155114},
  opturl = {https://link.aps.org/doi/10.1103/PhysRevB.90.155114}
}

@article{armitage10,
  title = {Progress and perspectives on electron-doped cuprates},
  author = {Armitage, N. P. and Fournier, P. and Greene, R. L.},
  journal = {Rev. Mod. Phys.},
  volume = {82},
  issue = {3},
  pages = {2421--2487},
  numpages = {0},
  year = {2010},
  month = {Sep},
  publisher = {American Physical Society},
  opdoi = {10.1103/RevModPhys.82.2421},
  opurl = {https://link.aps.org/doi/10.1103/RevModPhys.82.2421}
}

@article{bejas12,
  title = {Possible charge instabilities in two-dimensional doped {Mott} insulators},
  author = {Bejas, Mat\'{\i}as and Greco, Andr\'es and Yamase, Hiroyuki},
  journal = {Phys. Rev. B},
  volume = {86},
  issue = {22},
  pages = {224509},
  numpages = {12},
  year = {2012},
  month = {Dec},
  publisher = {American Physical Society},
  optdoi = {10.1103/PhysRevB.86.224509},
  opturl = {https://link.aps.org/doi/10.1103/PhysRevB.86.224509}}

@article{damascelli03,
  title = {Angle-resolved photoemission studies of the cuprate superconductors},
  author = {Damascelli, Andrea and Hussain, Zahid and Shen, Zhi-Xun},
  journal = {Rev. Mod. Phys.},
  volume = {75},
  issue = {2},
  pages = {473--541},
  numpages = {0},
  year = {2003},
  month = {Apr},
  publisher = {American Physical Society},
  doi = {10.1103/RevModPhys.75.473},
  url = {https://link.aps.org/doi/10.1103/RevModPhys.75.473}
}

@article {da-silva-neto15,
	author = {da Silva Neto, Eduardo H. and Comin, Riccardo and He, Feizhou and Sutarto, Ronny and Jiang, Yeping and Greene, Richard L. and Sawatzky, George A. and Damascelli, Andrea},
	title = {{Charge ordering in the electron-doped superconductor Nd$_{2-x}$Ce$_x$CuO$_4$}},
	volume = {347},
	number = {6219},
	pages = {282--285},
	year = {2015},
	doi = {10.1126/science.1256441},
	publisher = {American Association for the Advancement of Science},
	OPTissn = {0036-8075},
	optURL = {http://science.sciencemag.org/content/347/6219/282},
	OPTeprint = {http://science.sciencemag.org/content/347/6219/282.full.pdf},
	journal = {Science}
}

@article{da-silva-neto16,
	author = {da Silva Neto, Eduardo H. and Yu, Biqiong and Minola, Matteo and Sutarto, Ronny and Schierle, Enrico and Boschini, Fabio and Zonno, Marta and Bluschke, Martin and Higgins, Joshua and Li, Yangmu and Yu, Guichuan and Weschke, Eugen and He, Feizhou and Le Tacon, Mathieu and Greene, Richard L. and Greven, Martin and Sawatzky, George A. and Keimer, Bernhard and Damascelli, Andrea},
	title = {Doping-dependent charge order correlations in electron-doped cuprates},
	volume = {2},
	number = {8},
	pages = {e1600782},
	year = {2016},
	doi = {10.1126/sciadv.1600782},
	publisher = {American Association for the Advancement of Science},
	optURL = {http://advances.sciencemag.org/content/2/8/e1600782},
	OPTeprint = {http://advances.sciencemag.org/content/2/8/e1600782.full.pdf},
	journal = {Science Advances}}

@article{da-silva-neto18,
  title = {Coupling between dynamic magnetic and charge-order correlations in the cuprate superconductor {${\mathrm{Nd}}_{2\ensuremath{-}x}{\mathrm{Ce}}_{x}{\mathrm{CuO}}_{4}$}},
  author = {da Silva Neto, E. H. and Minola, M. and Yu, B. and Tabis, W. and Bluschke, M. and Unruh, D. and Suzuki, H. and Li, Y. and Yu, G. and Betto, D. and Kummer, K. and Yakhou, F. and Brookes, N. B. and Le Tacon, M. and Greven, M. and Keimer, B. and Damascelli, A.},
  journal = {Phys. Rev. B},
  volume = {98},
  issue = {16},
  pages = {161114},
  numpages = {6},
  year = {2018},
  month = {Oct},
  publisher = {American Physical Society},
  doi = {10.1103/PhysRevB.98.161114},
  url = {https://link.aps.org/doi/10.1103/PhysRevB.98.161114}
}

@article{kuboki17,
   author = {Kuboki, Kazuhiro and Yamase, Hiroyuki},
   title = {Static spin susceptibility in magnetically ordered states},
   journal = {Phys. Rev. B},
   volume = {96},
   number = {6},
   pages = {064411},
   ISSN = {2469-9950
2469-9969},
   DOI = {10.1103/PhysRevB.96.064411},
   year = {2017},
   type = {Journal Article}
}

@article {kunisada20,
	author = {Kunisada, So and Isono, Shunsuke and Kohama, Yoshimitsu and Sakai, Shiro and Bareille, C{\'e}dric and Sakuragi, Shunsuke and Noguchi, Ryo and Kurokawa, Kifu and Kuroda, Kenta and Ishida, Yukiaki and Adachi, Shintaro and Sekine, Ryotaro and Kim, Timur K. and Cacho, Cephise and Shin, Shik and Tohyama, Takami and Tokiwa, Kazuyasu and Kondo, Takeshi},
	title = {{Observation of small Fermi pockets protected by clean CuO2 sheets of a high-Tc superconductor}},
	volume = {369},
	number = {6505},
	pages = {833--838},
	year = {2020},
	optdoi = {10.1126/science.aay7311},
	publisher = {American Association for the Advancement of Science},
	optissn = {0036-8075},
	optURL = {https://science.sciencemag.org/content/369/6505/833},
	opteprint = {https://science.sciencemag.org/content/369/6505/833.full.pdf},
	journal = {Science}
}

@article{lee06,
  title = {Doping a {Mott} insulator: Physics of high-temperature superconductivity},
  author = {Lee, Patrick A. and Nagaosa, Naoto and Wen, Xiao-Gang},
  journal = {Rev. Mod. Phys.},
  volume = {78},
  issue = {1},
  pages = {17--85},
  numpages = {0},
  year = {2006},
  month = {Jan},
  publisher = {American Physical Society},
  optdoi = {10.1103/RevModPhys.78.17},
  opturl = {https://link.aps.org/doi/10.1103/RevModPhys.78.17}
}

@article{li17,
  title = {{Nature of the effective interaction in electron-doped cuprate superconductors: A sign-problem-free quantum Monte Carlo study}},
  author = {Li, Zi-Xiang and Wang, Fa and Yao, Hong and Lee, Dung-Hai},
  journal = {Phys. Rev. B},
  volume = {95},
  issue = {21},
  pages = {214505},
  numpages = {7},
  year = {2017},
  month = {Jun},
  publisher = {American Physical Society},
  doi = {10.1103/PhysRevB.95.214505},
  url = {https://link.aps.org/doi/10.1103/PhysRevB.95.214505}
}

@article{liu11,
	Author = {{R. H. Liu, X. G. Luo, M. Zhang, A. F. Wang, J. J. Ying, X. F. Wang, Y. J. Yan, Z. J. Xiang, P. Cheng, G. J. Ye, Z. Y. Li, and X. H. Chen}},
	Title = {{Coexistence of Superconductivity and Antiferromagnetism in Single Crystals $A_{0.8}$Fe$_{2-y}$Se$_2$ ($A$ = K, Rb, Cs, Tl/K and Tl/Rb): Evidence from Magnetization and Resistivity}},
	Journal = {Europhys. Lett.},
	Pages = {27008},
	Volume = {94},
	Year = {2011}}

@article{mukuda12,
	Author = {H. Mukuda and S. Shimizu and A. Iyo and Y. Kitaoka},
	Title = {{High-$T_c$ Superconductivity and Antiferromagnetism in Multilayered Copper Oxides---A New Paradigm of Superconducting Mechanism---}},
	Journal = {J. Phys. Soc. Jpn.},
	Optauthor = {si and {\it et al.}},
	Pages = {011008},
	Volume = {81},
	Year = {2012}}

@article{pratt09,
	Author = {{D. K. Pratt, W. Tian, A. Kreyssig, J. L. Zarestky, S. Nandi, N. Ni, S. L. Bud'ko, P. C. Canfield, A. I. Goldman, and R. J. McQueeney}},
	Title = {{Coexistence of Competing Antiferromagnetic and Superconducting Phases in the Underdoped Ba(Fe$_{0.953}$Co$_{0.047}$)$_2$As$_2$ Compound Using X-Ray and Neutron Scattering Techniques}},
	Journal = {Phys.\ Rev.\ Lett.},
	Pages = {087001},
	Volume = {103},
	Year = {2009}}

@article{sachdev13,
  title = {Bond Order in Two-Dimensional Metals with Antiferromagnetic Exchange Interactions},
  author = {Sachdev, Subir and La Placa, Rolando},
  journal = {Phys. Rev. Lett.},
  volume = {111},
  issue = {2},
  pages = {027202},
  numpages = {5},
  year = {2013},
  month = {Jul},
  publisher = {American Physical Society},
  optdoi = {10.1103/PhysRevLett.111.027202},
  opturl = {https://link.aps.org/doi/10.1103/PhysRevLett.111.027202}
}

@article{yamase15b,
	author = {Yamase, Hiroyuki and Bejas, Mat\'{\i}as and Greco, Andr\'es},
	title = {d-wave bond-order charge excitations in electron-doped cuprates},
	optDOI= "10.1209/0295-5075/111/57005",
	opturl= "https://doi.org/10.1209/0295-5075/111/57005",
	journal = {Europhys. Lett.},
	year = {2015},
	volume = {111},
	number = {5},
	pages = {57005}}

@article{yamase19, 
  title = {Doping dependence of $d$-wave bond-charge excitations in electron-doped cuprates},
  author = {Yamase, Hiroyuki and Bejas, Mat\'{\i}as and Greco, Andr\'es},
  journal = {Phys. Rev. B},
  volume = {99},
  issue = {1},
  pages = {014513},
  numpages = {7},
  year = {2019},
  month = {Jan},
  publisher = {American Physical Society},
  optdoi = {10.1103/PhysRevB.99.014513},
  opturl = {https://link.aps.org/doi/10.1103/PhysRevB.99.014513}
}

@article{yamase21b,
  title = {Anomaly of longitudinal spin susceptibility at a superconducting instability inside a magnetic phase},
  author = {Yamase, Hiroyuki and Zafur, Muhammad},
  journal = {Phys. Rev. B},
  volume = {103},
  issue = {22},
  pages = {224527},
  numpages = {9},
  year = {2021},
  month = {Jun},
  publisher = {American Physical Society},
  optdoi = {10.1103/PhysRevB.103.224527},
  opturl = {https://link.aps.org/doi/10.1103/PhysRevB.103.224527}
}

@article{yamase21c,
author = {Yamase ,Hiroyuki},
title = {Theoretical Insights into Electronic Nematic Order, Bond-Charge Orders, and Plasmons in Cuprate Superconductors},
journal = {J. Phys. Soc. Jpn.},
volume = {90},
number = {11},
pages = {111011},
year = {2021},
optdoi = {10.7566/JPSJ.90.111011},
optURL = {https://doi.org/10.7566/JPSJ.90.111011},
opteprint = {https://doi.org/10.7566/JPSJ.90.111011}
}

@article{zafur24,
  title = {Spin and bond-charge excitation spectra in correlated electron systems near an antiferromagnetic phase},
  author = {Zafur, Muhammad and Yamase, Hiroyuki},
  journal = {Phys. Rev. B},
  volume = {109},
  issue = {24},
  pages = {245127},
  numpages = {26},
  year = {2024},
  month = {Jun},
  publisher = {American Physical Society},
  doi = {10.1103/PhysRevB.109.245127},
  url = {https://link.aps.org/doi/10.1103/PhysRevB.109.245127}
}
\end{document}